\documentclass[sigconf,authorversion,nonacm]{acmart}

\usepackage{etex} %

\usepackage{siunitx}
\usepackage{verbatim}
\usepackage{graphicx,afterpage}
\usepackage{yfonts}
\usepackage{amsmath}
\usepackage{mathtools}
\usepackage{stmaryrd}
\usepackage{amsfonts}
\usepackage{calc}
\usepackage{xspace}
\usepackage{listings}
\usepackage{multirow, booktabs}
\usepackage{wrapfig}
\usepackage{enumitem}
\usepackage{caption}
\usepackage[labelformat=simple]{subcaption}
\usepackage{dblfloatfix} %

\usepackage{fancyhdr}

\usepackage[T1]{fontenc} %

\usepackage{array} %
\usepackage{rotating}
\lstdefinestyle{custompython}{
 belowskip=0in,
 belowcaptionskip=-10pt,
 breaklines=true,
 captionpos=b,
 language=Python,
 showstringspaces=false,
 numbers=left,
 stepnumber=1,
 basicstyle={\linespread{0.8}\fontseries{sb}\small\ttfamily},
 keywordstyle=\bfseries,
 xleftmargin=2em,
 frame=single,
 framexleftmargin=2em,
 commentstyle=\itshape\color{green!40!black},
 morekeywords={to,yield},
}

\definecolor{ltred}{HTML}{FB9A99}
\definecolor{dkred}{HTML}{E41A1C}
\definecolor{dkorange}{HTML}{F07F03}
\definecolor{ltorange}{HTML}{F7C090}
\definecolor{dkgreen}{HTML}{3BA32F}
\definecolor{ltgreen}{HTML}{B2E089}
\definecolor{dkblue}{HTML}{2879B4}
\definecolor{ltblue}{HTML}{A6CEE2}
\definecolor{dkpurple}{HTML}{6A3D9A}
\definecolor{ltpurple}{HTML}{CAB1D7}

\usepackage{hyphenat}
\setlist[enumerate]{leftmargin=0.25in,itemsep=1pt,topsep=1pt}

\makeatletter
\renewcommand{\paragraph}[1]{\noindent {\bf #1}}
\renewcommand{\subparagraph}[1]{\noindent {\emph{\textbf{#1}}}}
\makeatother

\definecolor{lcolor}{RGB}{0, 56, 186} %
\usepackage{color}
\definecolor{tableaublue}{rgb}{0.44,0.62,0.81}
\definecolor{tableauorange}{rgb}{0.9,0.55,0.25} %
\definecolor{tableaugreen}{rgb}{0.4,0.75,0.36}
\definecolor{tableaured}{rgb}{0.93,0.4,0.36}
\definecolor{tableaupurple}{rgb}{0.68,0.55,0.79}

\usepackage{pifont}
\usepackage{fourier-orns}
\definecolor{darkred}{rgb}{.65,0,0}
\definecolor{darkgreen}{rgb}{0,.5,0}
\definecolor{darkyellow}{rgb}{0.95,.6,0.1}

\usepackage{textcomp}
\newcommand{\name}{F1\xspace}

\renewcommand{\thefootnote}{\fnsymbol{footnote}}

\usepackage{comment}

\newcommand{\figArch}{
  \begin{figure}[t]
        \begin{center}
     \includegraphics[width=\columnwidth]{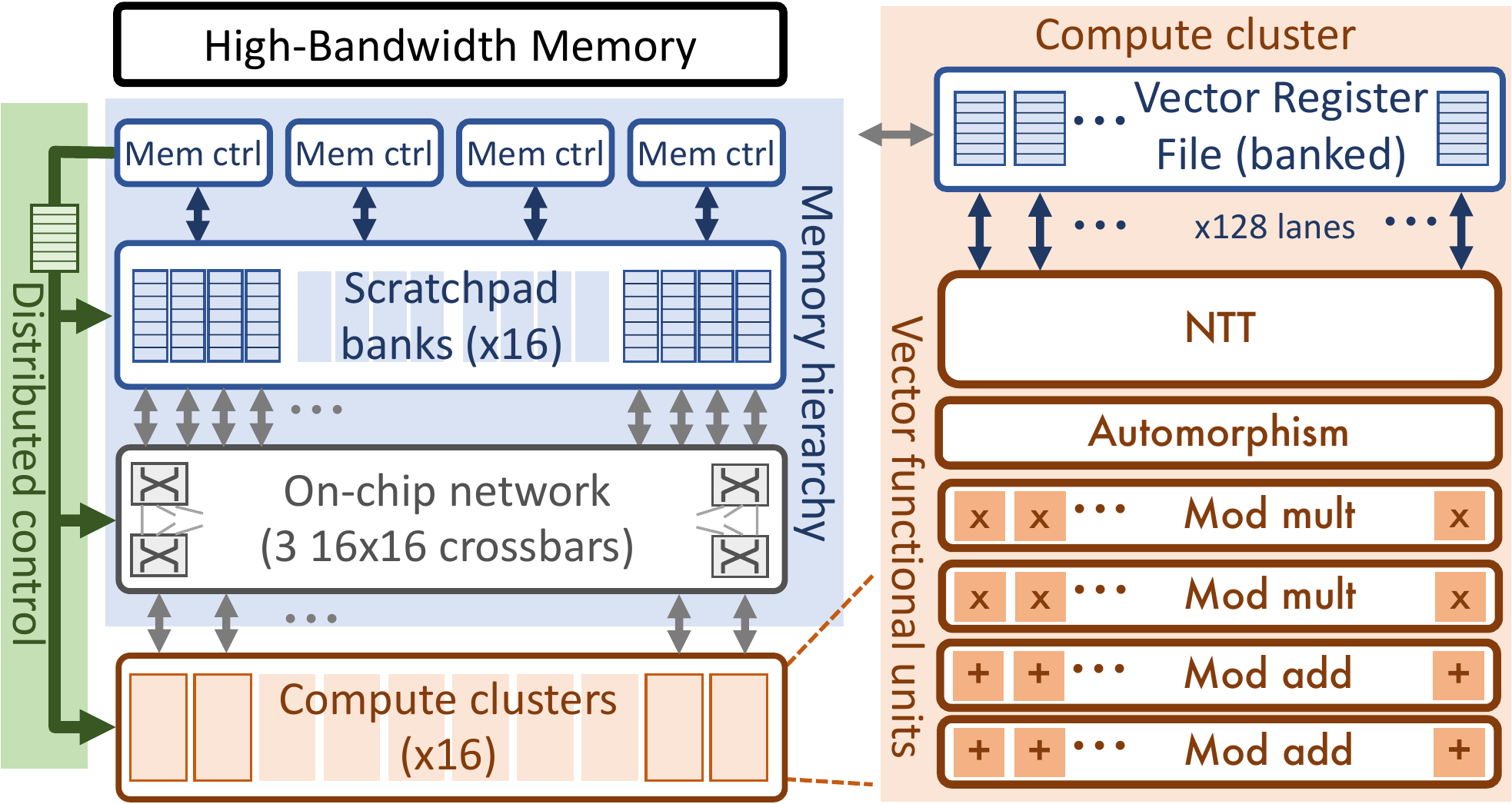}
     \caption{Overview of the \name architecture.}
         \vspace{0.2cm}
    \label{fig:arch}
  \end{center}
  \end{figure}
}

\newcommand{\figMultDataflow}{
  \begin{figure}[t]
        \begin{center}
     \includegraphics[width=0.99\columnwidth]{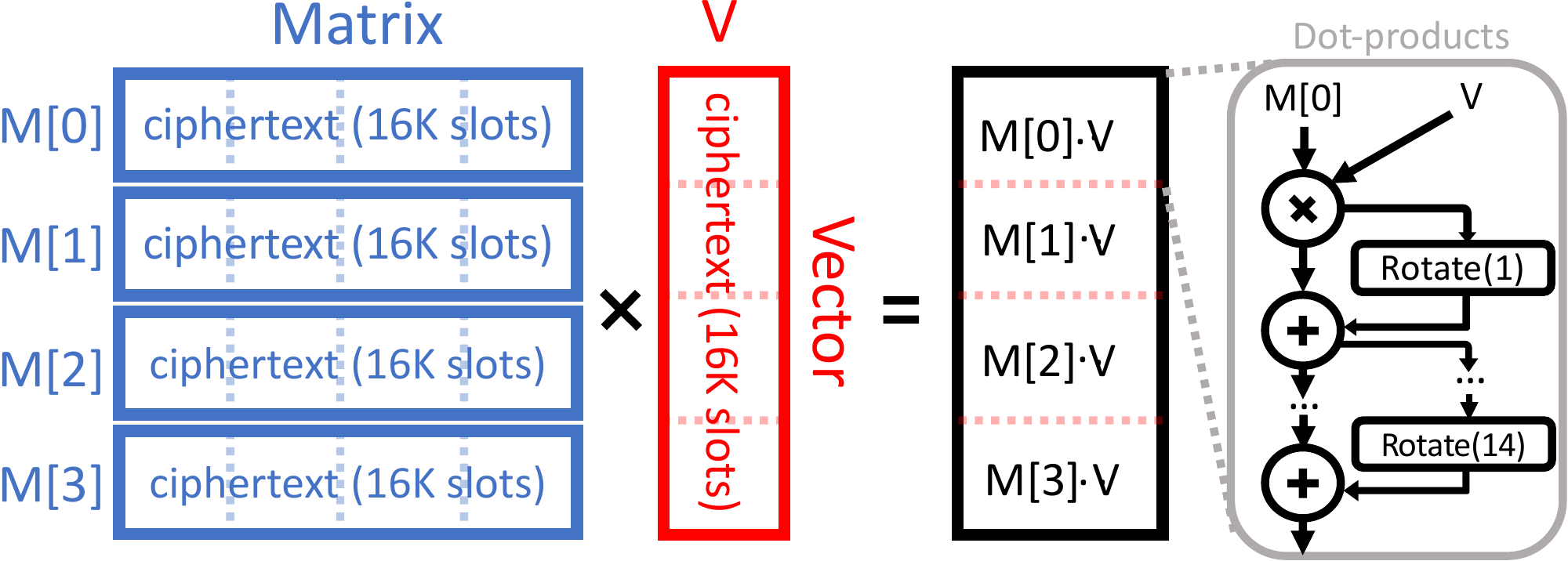}
    \caption{Example matrix-vector multiply using FHE.}
    \label{fig:MultDataflow}
    \vspace{0.2cm}
    \end{center}
  \end{figure}
}

\newcommand{\figOpBreakdown}{
    \begin{figure}[t]
    \begin{center}
        \includegraphics[width=0.99\columnwidth]{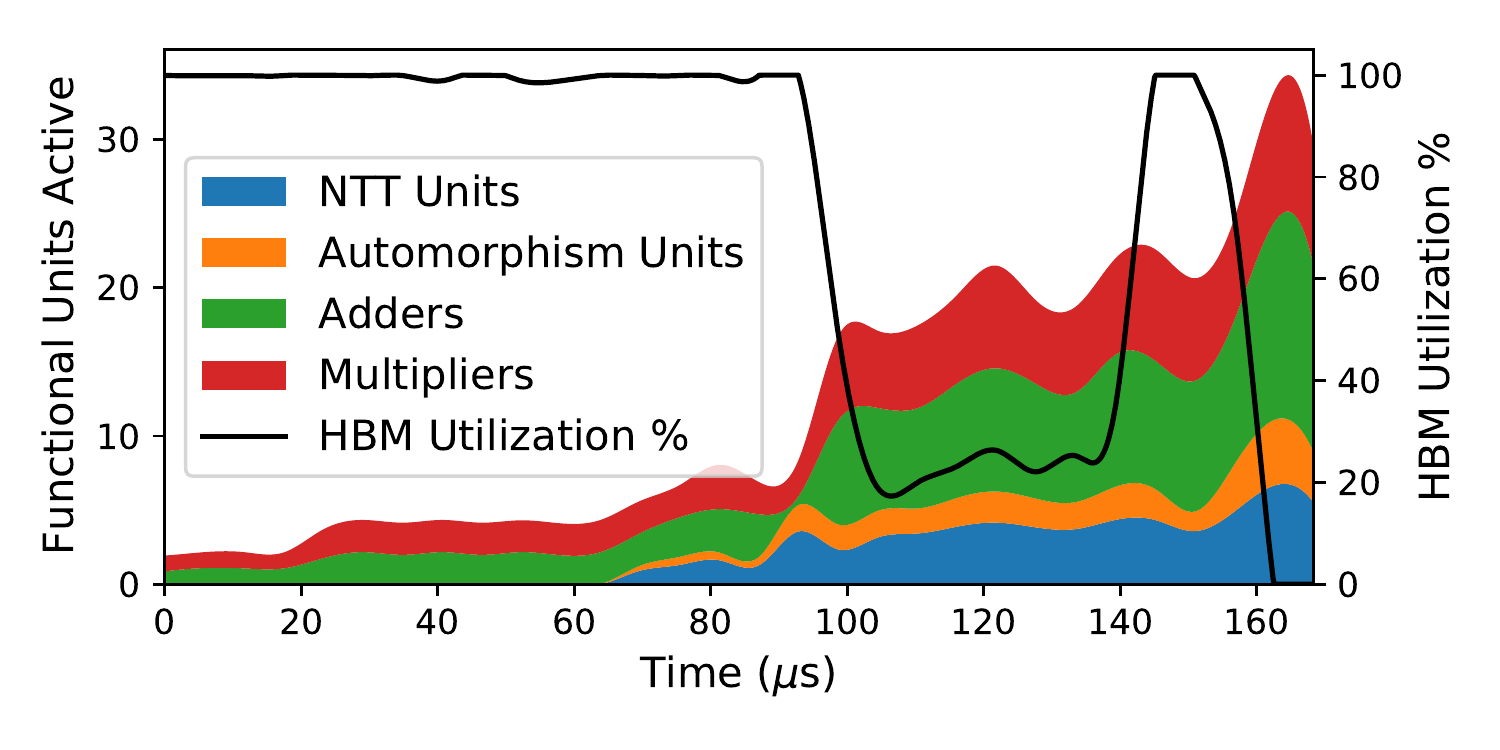}
        \vspace{-0.05in}
        \caption{Functional unit and HBM utilization over time for the LoLa-MNIST PTW benchmark.}
        \vspace{0.2cm}
        \label{fig:opBreakdown}
    \end{center}
    \end{figure}
}

\newcommand{\figCompilerOverview}{
  \begin{figure}[t]
        \begin{center}
     \includegraphics[width=\columnwidth]{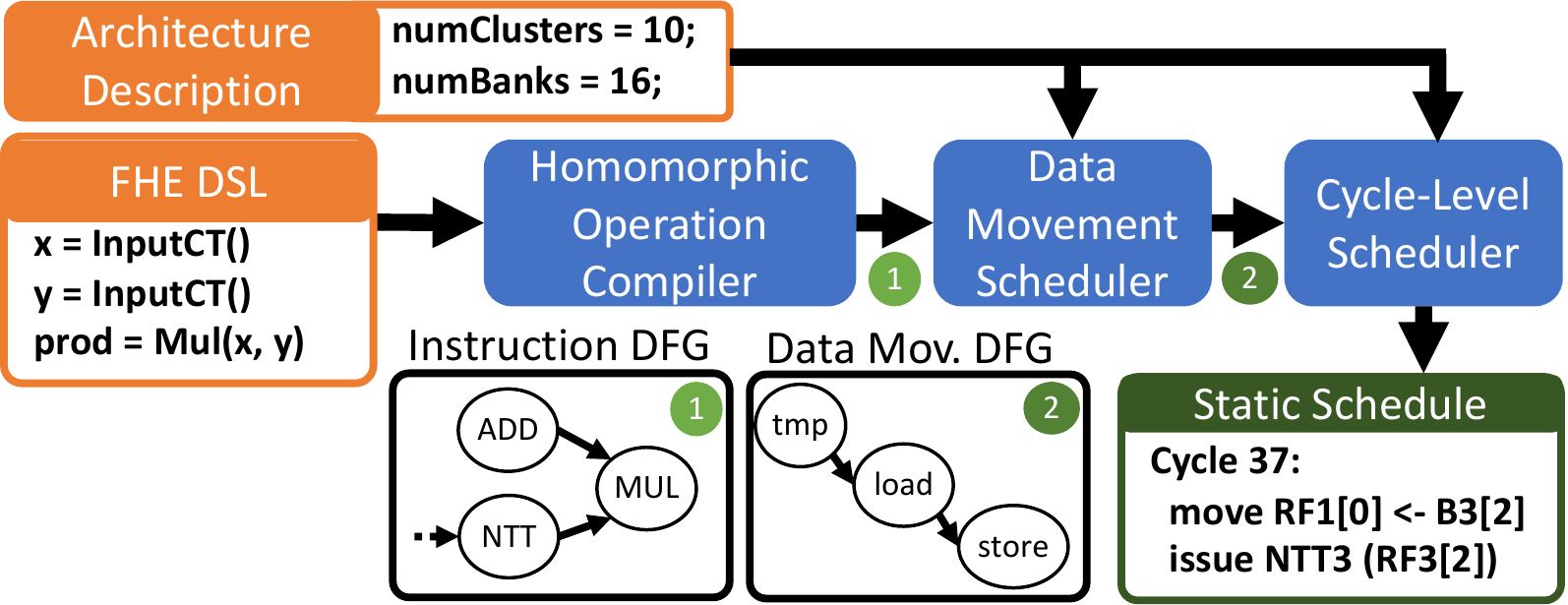}
    \caption{Overview of the \name compiler.}
    \label{fig:compilerOverview}
    \vspace{0.2cm}    
    \end{center}
  \end{figure}
}

\newcommand{\figautfu}{
\setlength{\columnsep}{7pt}
  \begin{wrapfigure}{r}{0.22\linewidth}
      \vspace{-1.8em}
     \includegraphics[width=0.25\columnwidth]{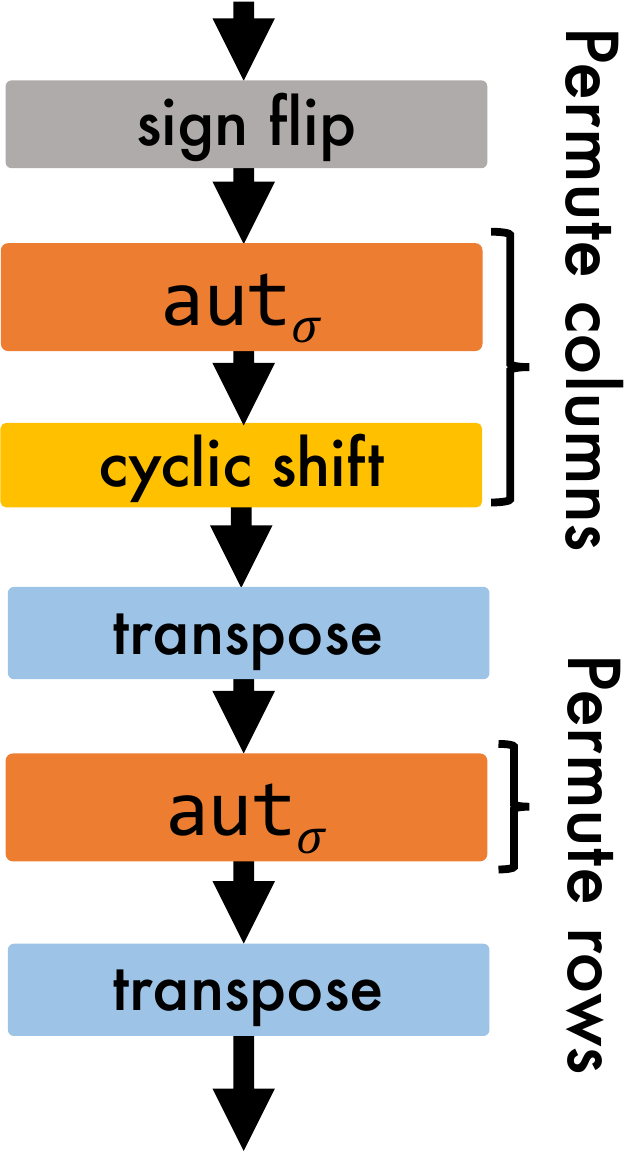}
    \caption{Automorphism unit.}
    \label{fig:aut_fu}
  \end{wrapfigure}
}

\newcommand{\figAutomorphism}{
  \begin{figure}[t]
        \begin{center}
    \includegraphics[width=0.99\columnwidth]{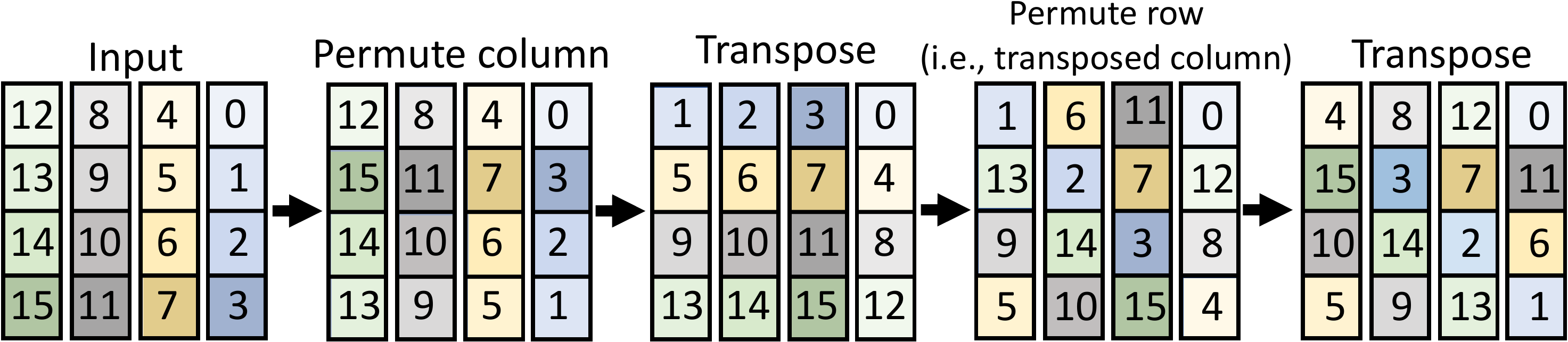}
    \caption{Applying $\sigma_3$ on an RNS polynomial of four 4-element chunks by using only permutations local to chunks.}
    \vspace{0.2cm}
    \label{fig:automorphism}
    \end{center}
  \end{figure}
}

\newcommand{\figFourStepNTT}{
  \begin{figure}[t]
    \includegraphics[width=0.99\columnwidth]{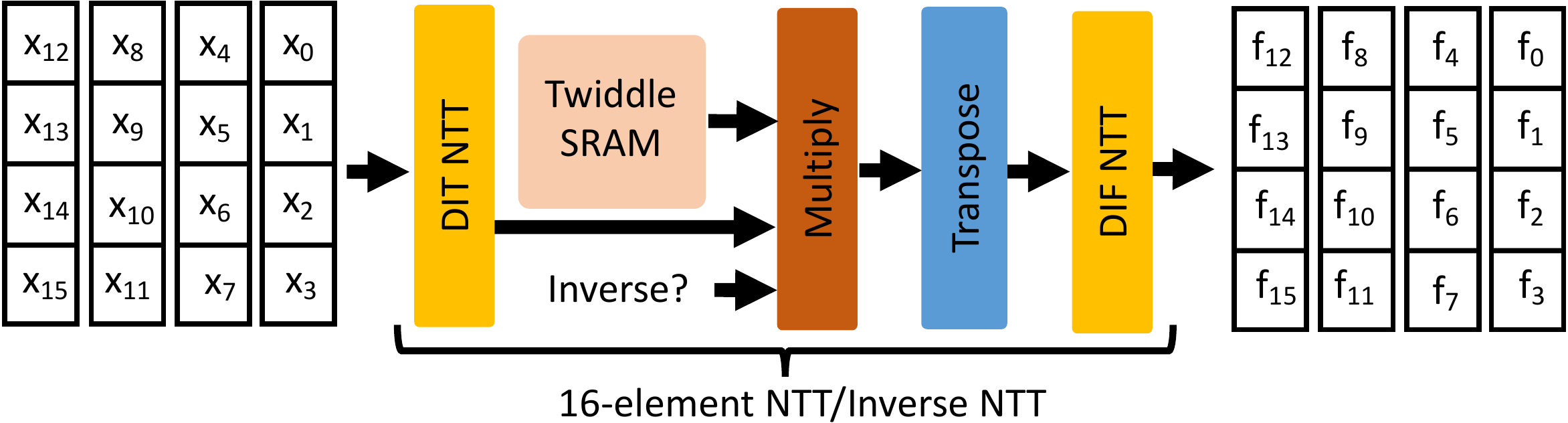}
    \caption{Example of a four-step NTT datapath that uses 4-point NTTs to implement 16-point NTTs.}
    \label{fig:fourStepNTT}
    \vspace{0.2cm}
  \end{figure}
}

\newcommand{\figQuadrantSwap}{
  \begin{figure}[t]
    \centering
    \includegraphics[width=0.99\columnwidth]{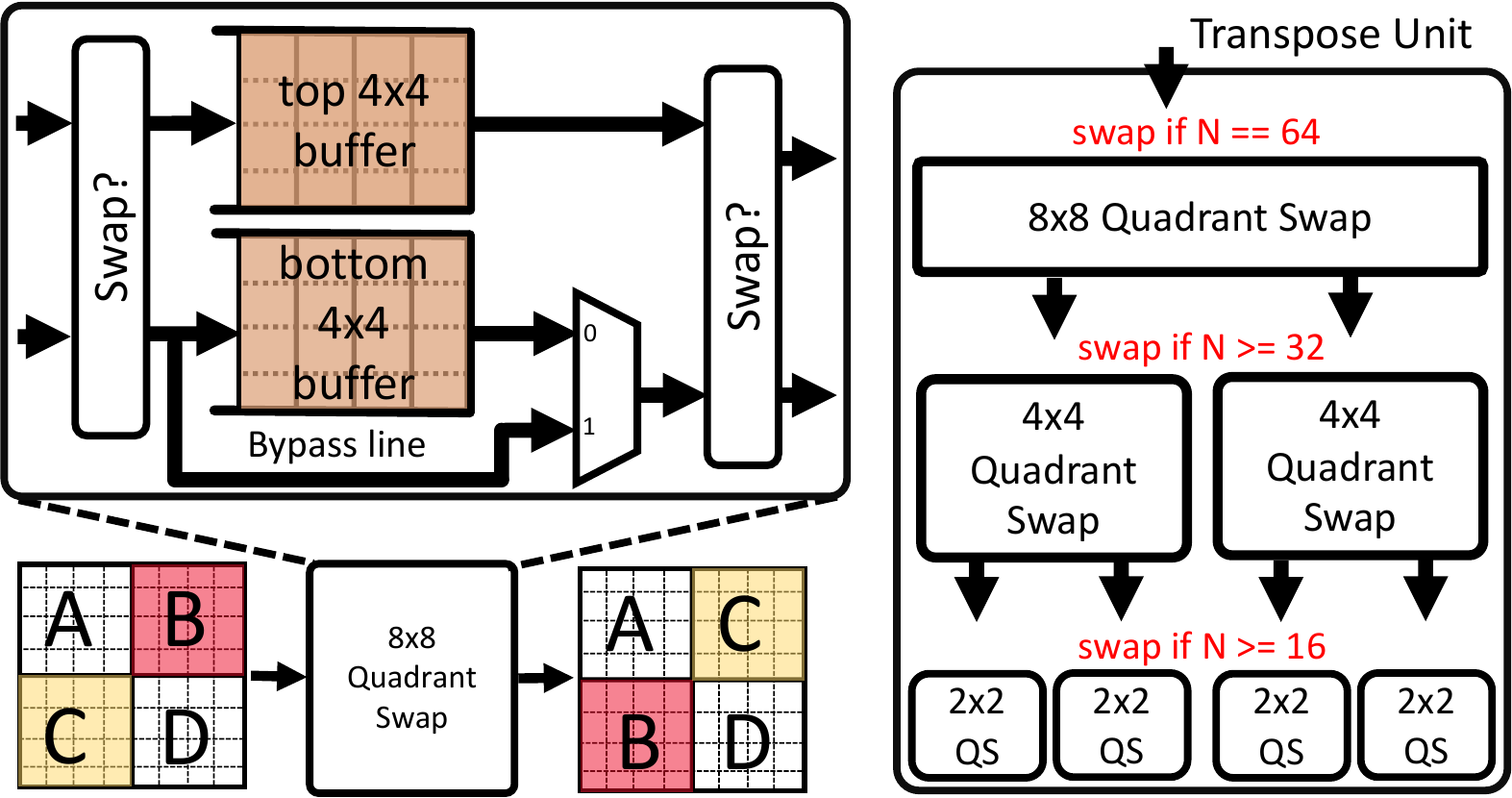}
    \caption{Transpose unit (right) and its component quadrant-swap unit (left).}
    \label{fig:quadrantSwap}
    \vspace{0.2cm}
  \end{figure}
}

\newcommand{\figOverview}{
  \begin{figure}[t]
    \centering
    \includegraphics[width=\columnwidth]{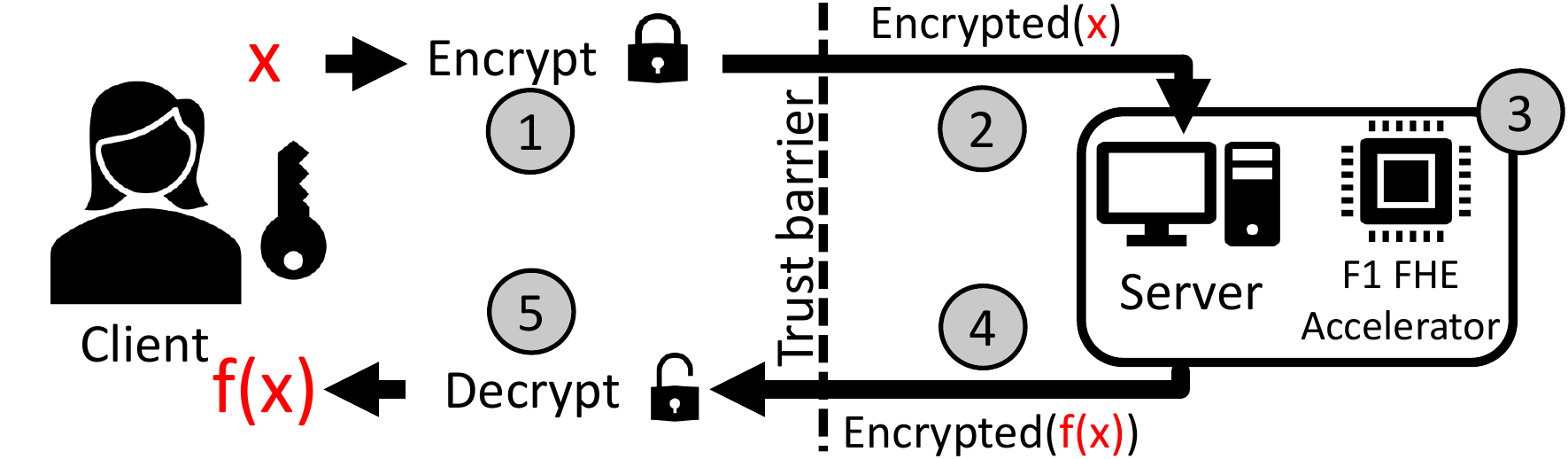}
    \caption{FHE allows a user to securely offload computation to an untrusted server.}
    \label{fig:overview}
    \vspace{0.2cm}
  \end{figure}
}

\newcommand{\figDataMovement}{
\begin{figure}
  \centering
  \captionsetup[subfigure]{labelformat=empty}
  \begin{subfigure}[b]{0.49\columnwidth}
    \includegraphics[width=\linewidth]{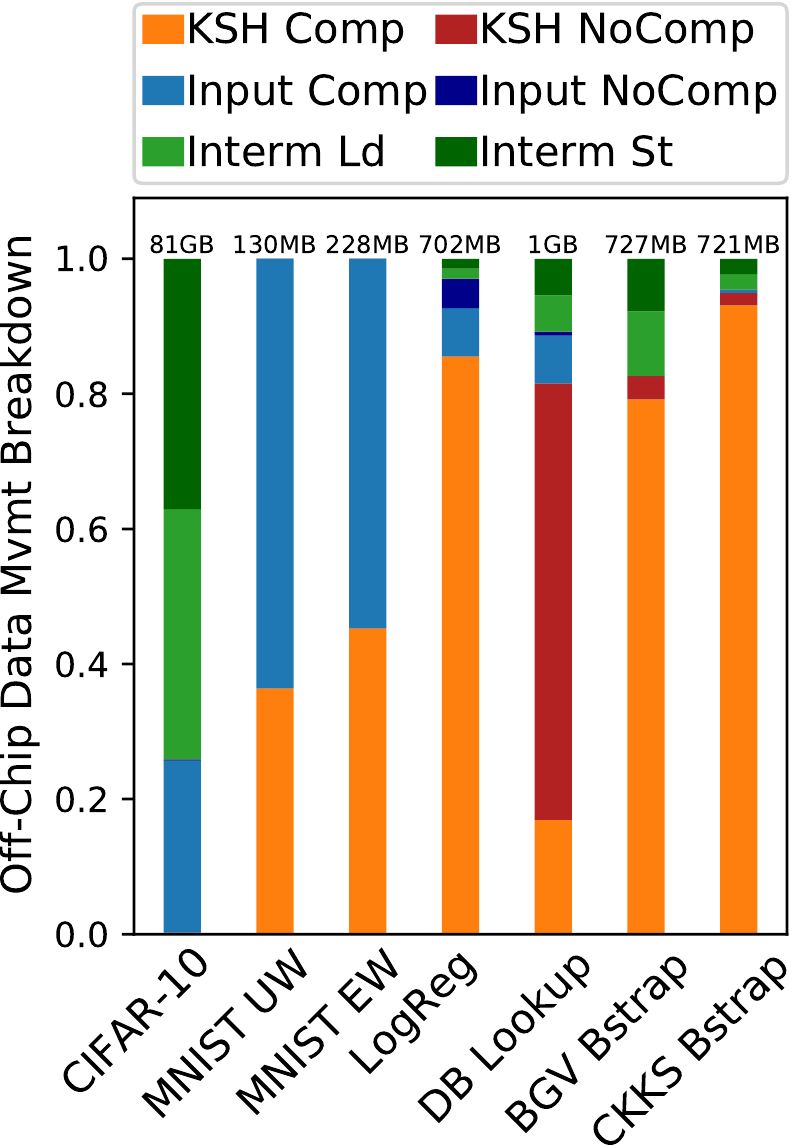}
    \caption{}
    \label{fig:dataMovement}
  \end{subfigure}
  \hfill %
  \begin{subfigure}[b]{0.49\columnwidth}
    \includegraphics[width=\linewidth]{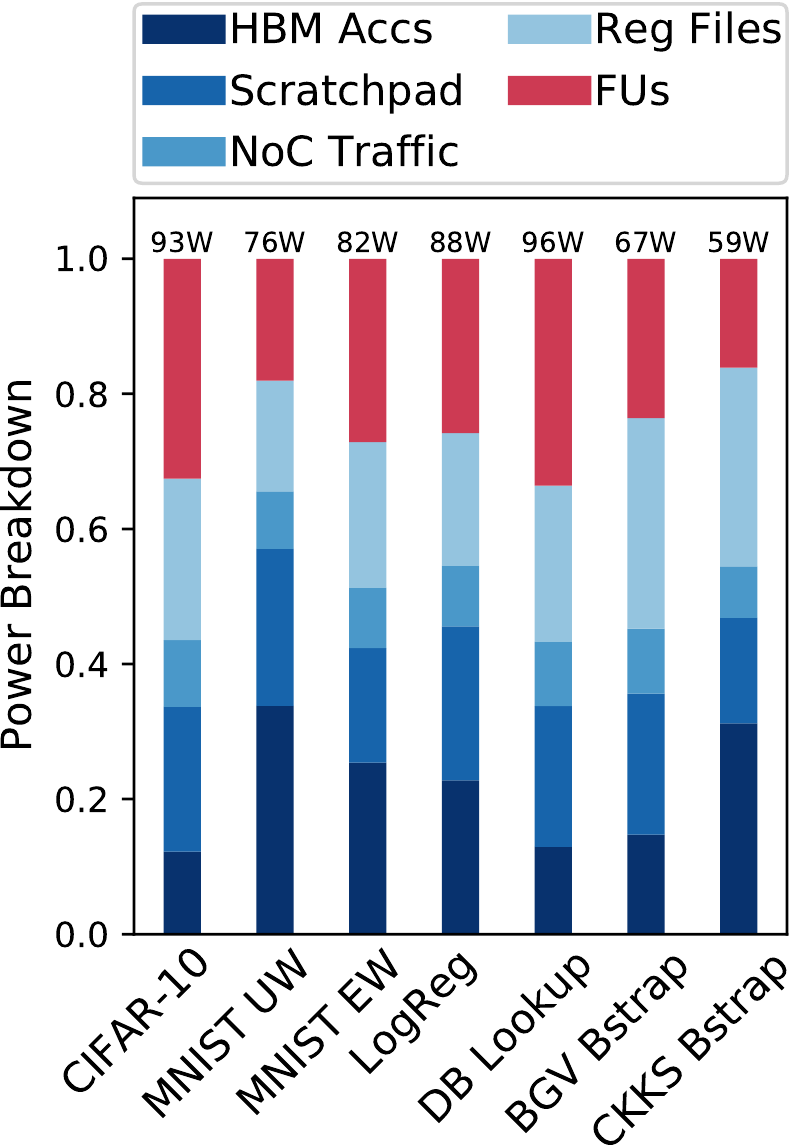}
    \caption{}
    \label{fig:power}
  \end{subfigure}
  \vspace*{-0.1in}
  \begin{center}
    (a) \hspace{0.45\columnwidth} (b)
  \end{center}
  \caption{Per-benchmark breakdowns of (a) data movement and (b) average power for \name.}
  \vspace{0.2cm}
\end{figure}
}

\newcommand{\figConfigs}{
    \setlength{\columnsep}{7pt}
  \begin{wrapfigure}{t}{0.5\linewidth}
    \begin{center}
    \vspace{-0.3in}
     \includegraphics[width=0.5\columnwidth]{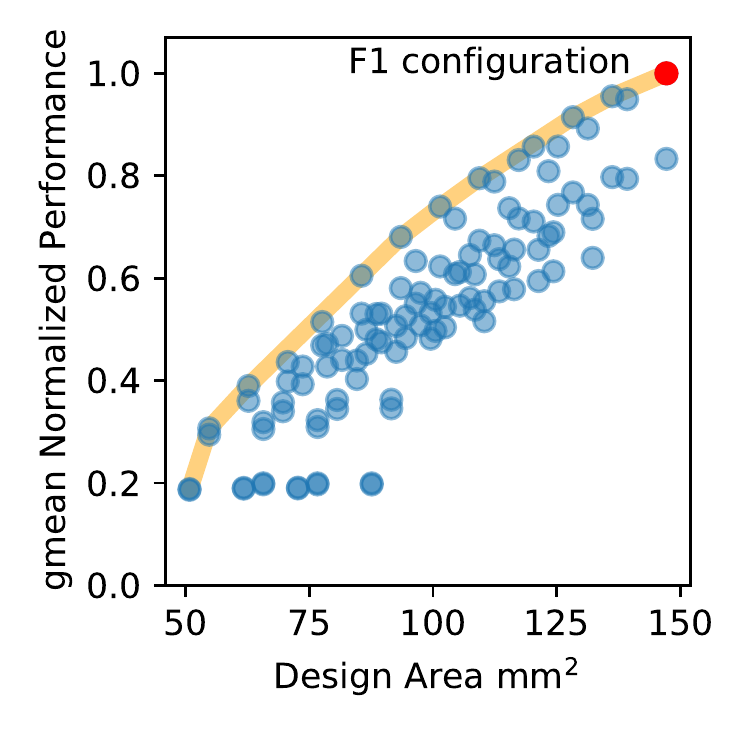}
    \caption{Performance vs. area across \name configurations.}
    \label{fig:pareto}
    \end{center}
  \end{wrapfigure}
}

\newcommand{\tblGF}{
  \begin{table}[t]
    \begin{center}
      \begin{normalsize}
        \begin{tabular}{lrrr}
          \toprule
          Component & Area [mm$^2$] & TDP [W] \\
          \midrule
          NTT FU & 2.27 & 4.80  \\
          Automorphism FU & 0.58 & 0.99  \\
          Multiply FU & 0.25 & 0.60  \\
          Add FU & 0.03 & 0.05  \\
          Vector RegFile (512\,KB) & 0.56 & 1.67 \\
          \textbf{Compute cluster} & 3.97 & 8.75 \\
          (NTT, Aut, 2$\times$ Mul, 2$\times$ Add, RF) & & \\
          \textbf{Total compute} (16 clusters) &  \textbf{63.52} & \textbf{140.0} \\
          \midrule
          Scratchpad (16$\times$4\,MB banks) & 48.09 & 20.35 \\
          3$\times$NoC (16$\times$16 512\,B bit-sliced~\cite{passas:tocaid12:crossbar}) & 10.02 & 19.65 \\
          Memory interface (2$\times$HBM2 PHYs) & 29.80 & 0.45 \\
          \textbf{Total memory system} &  \textbf{87.91} & \textbf{40.45} \\
          \midrule
          \textbf{Total \name} &  \textbf{151.4} & \textbf{180.4} \\
          \bottomrule
        \end{tabular}
      \end{normalsize}
    \end{center}
    \caption{Area and Thermal Design Power (TDP) of \name, and breakdown by component.}
    \label{tbl:GF12}
  \end{table}
}

\newcommand{\tblModMult}{
  \begin{table}[b]
    \begin{normalsize}
      \begin{center}
        \begin{tabular}{lrrr}
          \toprule
          Multiplier & Area [$\mu$m$^2$] & Power [mW] & Delay [ps] \\
          \midrule
          Barrett & $5,271$ & $18.40$ & 1,317 \\
          Montgomery & $2,916$ & $9.29$ & 1,040 \\
          NTT-friendly & $2,165$ & $5.36$ & 1,000 \\ %
          \midrule
          \textbf{FHE-friendly (ours)} & $1,817$ & $4.10$ & 1,000 \\
          \bottomrule
        \end{tabular}
        \caption{Area, power, and delay of modular multipliers.}
        \label{tbl:modMult}
        \vspace{-0.2in}
      \end{center}
    \end{normalsize}
  \end{table}
}

\newcommand{\x}{$\times$}

\newcommand{\tblMicrobenchmark}{
  \begin{table*}[b]
    \begin{normalsize}
      \begin{center}
      \begin{tabular}{l|rrr|rrr|rrr}
        \toprule
        & \multicolumn{3}{c|}{$N = 2^{12}$, $\log Q = 109$} &
        \multicolumn{3}{c|}{$N = 2^{13}$, $\log Q = 218$} &
        \multicolumn{3}{c}{$N = 2^{14}$, $\log Q = 438$} \\
        
        & \textbf{\name} & vs.\ CPU & vs.\ HEAX$_\sigma$ & \textbf{\name} & vs.\ CPU & vs.\ HEAX$_\sigma$ & \textbf{\name} & vs.\ CPU & vs.\ HEAX$_\sigma$\\

        \midrule

        NTT
        &\textbf{12.8} %
        &17,148$\times$ %
        &1,600$\times$ %

        &\textbf{44.8} %
        &10,736$\times$ %
        &1,733$\times$ %

        &\textbf{179.2} %
        &8,838$\times$ %
        &1,866$\times$ %
        \\
        Automorphism 
        &\textbf{12.8} %
        &7,364$\times$ %
        &440$\times$ %

        &\textbf{44.8} %
        &8,250$\times$ %
        &426$\times$ %

        &\textbf{179.2} %
        &16,957$\times$ %
        &430$\times$ %
        \\

        \midrule

        Homomorphic multiply
        &\textbf{60.0} %
        &48,640$\times$ %
        &172$\times$ %

        &\textbf{300} %
        &27,069$\times$ %
        &148$\times$ %

        &\textbf{2,000} %
        &14,396$\times$ %
        &190$\times$ %
        \\

        Homomorphic permutation
        &\textbf{40.0} %
        &17,488$\times$ %
        &256$\times$ %

        &\textbf{224} %
        &10,814$\times$ %
        &198$\times$ %

        &\textbf{1,680} %
        &6,421$\times$ %
        &227$\times$ %
        \\
        
        \bottomrule
      \end{tabular}
      \end{center}
      \caption{Performance on microbenchmarks: \name's \textbf{reciprocal throughput, in nanoseconds per ciphertext operation} (lower is better) and speedups over CPU and HEAX$_\sigma$ (HEAX augmented with scalar automorphism units) (higher is better).}
      \label{tbl:microbenchmark}
    \end{normalsize}
  \end{table*}
}

\newcommand{\tblBenchmark}{
  \begin{table}[t]
    \begin{normalsize}
      \begin{center}
      \begin{tabular}{lrrr}
        \toprule
        Execution time (ms) on & CPU & \name & Speedup \\
        
        \midrule
        LoLa-CIFAR Unencryp. Wghts. & $1.2\times10^6$ & \textbf{241} & $5,011$\x \\
        LoLa-MNIST Unencryp. Wghts. & $2,960$ & \textbf{0.17} & $17,412$\x \\
        LoLa-MNIST Encryp. Wghts. & $5,431$ & \textbf{0.36} & $15,086$\x \\
        Logistic Regression & $8,300$ & \textbf{1.15} & $7,217$\x \\
        DB Lookup & $29,300$ & \textbf{4.36} & $6,722$\x \\
        BGV Bootstrapping & $4,390$ & \textbf{2.40} & $1,830$\x \\  %
        CKKS Bootstrapping & $1,554$ & \textbf{1.30} & $1,195$\x \\    %
        \midrule

        \textbf{gmean speedup} &&& $5,432$\x \\
        \bottomrule
      \end{tabular}
   \end{center}
      \hfill\footnotemark[1]{LoLa's release did not include MNIST with encrypted weights, so we reimplemented it in HELib.}\quad\mbox{}
    \end{normalsize}
      \caption{Performance of \name and CPU on full FHE benchmarks: execution times in milliseconds
      and \name's speedup.}      
      \label{tbl:benchmark}
      \vspace{-0.055in} %
  \end{table}
}

\newcommand{\tblSensitivity}{
  \begin{table}[t]
    \begin{normalsize}
      \begin{center}
        \begin{tabular}{lrrr}
            \toprule
            Benchmark & LT NTT & LT Aut & CSR \\
            \midrule
            LoLa-CIFAR Unencryp. Wghts. & 3.5\x & 12.1\x & ---\footnotemark[1] \\
            LoLa-MNIST Unencryp. Wghts. & 5.0\x & 4.2\x & 1.1\x \\
            LoLa-MNIST Encryp. Wghts. & 5.1\x & 11.9\x & 7.5\x \\
            Logistic Regression & 1.7\x & 2.3\x & 11.7\x \\
            DB Lookup & 2.8\x & 2.2\x & ---\footnotemark[1] \\
            BGV Bootstrapping & 1.5\x & 1.3\x & 5.0\x \\
            CKKS Bootstrapping & 1.1\x & 1.2\x & 2.7\x \\
            \midrule
            \textbf{gmean speedup} & 2.5\x & 3.6\x & 4.2\x \\
            \bottomrule
        \end{tabular}
      \end{center}
      \hfill\footnotemark[1]{CSR is intractable for this benchmark.}\quad\mbox{}
    \end{normalsize}
        \caption{Speedups of \name over alternate configurations: %
          LT NTT/Aut = Low-throughput NTT/Automorphism FUs; CSR = Code Scheduling to minimize Register Usage~\cite{goodman:ics1988:code}.}
        \label{tbl:sensitivity}
  \end{table}
}

\usepackage{fancyhdr}
\usepackage[normalem]{ulem}
\usepackage[keeplastbox]{flushend}
\newcommand{\MIT}{1}
\newcommand{\UM}{2}
\newcommand{\SRI}{3}

\title{\name: A Fast and Programmable Accelerator \\ for Fully Homomorphic Encryption (Extended Version)}

\date{}

\def\addAuthorsAndEqualContribFootnote{1}  %

\ifx\addAuthorsAndEqualContribFootnote\undefined
\else
 \author{Axel Feldmann$^\MIT$$^\ast$, Nikola Samardzic$^\MIT$$^\ast$, Aleksandar Krastev$^\MIT$, Srini Devadas$^\MIT$, \\Ron Dreslinski$^\UM$, Karim Eldefrawy$^\SRI$, Nicholas Genise$^\SRI$, Chris Peikert$^\UM$, Daniel Sanchez$^\MIT$}
 \affiliation{\vspace{1em}\begin{tabular}{cc}
     $^\MIT$ Massachusetts Institute of Technology & $^\UM$ University of Michigan \\
     \sf \{axelf, nsamar, alexalex, devadas, sanchez\}@csail.mit.edu & \sf \{dreslin, cpeikert\}@umich.edu  \\
   \end{tabular}
   \begin{tabular}{cc}
   $^\SRI$ SRI International  \\
   \{karim.eldefrawy, nicholas.genise\}@sri.com
   \end{tabular}
   \country{}}
\newcommand{\cleanauthors}{Axel Feldmann, Nikola Samardzic, Aleksandar Krastev, Srini Devadas, Ron Dreslinski, Christopher Peikert, Daniel Sanchez}

\fi

\setcopyright{none}

\begin{document}

\begin{abstract}

Fully Homomorphic Encryption (FHE) allows computing on encrypted data, enabling secure offloading of computation to untrusted servers.
Though it provides ideal security, FHE is expensive when executed in software, 4 to 5 orders of magnitude slower than computing on unencrypted data.
These overheads are a major barrier to FHE's widespread adoption.

We present \name, the first FHE accelerator that is programm\-able, i.e., capable of executing full FHE programs.
\name builds on an in-depth architectural analysis of the characteristics of FHE computations
that reveals acceleration opportunities.
\name is a wide-vector processor with novel functional units deeply specialized to FHE primitives, 
such as modular arithmetic, number-theoretic transforms, and structured permutations.
This organization provides so much compute throughput that data movement becomes the key bottleneck.
Thus, \name is primarily designed to minimize 
data movement.
Hardware provides an explicitly managed memory hierarchy and mechanisms to decouple data movement from execution.
A novel compiler leverages these mechanisms to maximize reuse and schedule off-chip and on-chip data movement.
 
We evaluate \name using cycle-accurate simulation and RTL synthesis.
\name is the first system to accelerate complete FHE programs,
and outperforms state-of-the-art software implementations by gmean 5,400$\times$ and by up to 17,000$\times$.
These speedups counter most of FHE's overheads and enable new applications, like real-time private deep learning in the cloud.

\end{abstract}

\maketitle

\thispagestyle{firstpage}
\pagestyle{plain}

\ifx\addAuthorsAndEqualContribFootnote\undefined
\else
\begingroup\renewcommand\thefootnote{\small{$\ast$}}
\footnotetext{\small{\textbf{A.\ Feldmann and N.\ Samardzic contributed equally to this work.} \\ [0.5ex] This is an extended version of a paper that will appear in the Proceedings
of the 54th Annual IEEE/ACM International Symposium on Microarchitecture (MICRO), 2021~\cite{feldmann:micro21:f1}.}}
\endgroup
\fi

\section{Introduction}
\label{sec:intro}

Despite massive efforts to improve the security of computer systems,
security breaches are only becoming more frequent and damaging,
as more sensitive data is processed in the cloud~\cite{malekos-smith:csis20:hidden-costs-cybercrime,ibm20:breach-cost-report}.
Current encryption technology is of limited help,
because servers must decrypt data before processing it.
Once data is decrypted, it is vulnerable to breaches.

Fully Homomorphic Encryption (FHE) is a class of encryption schemes that
address this problem by enabling \emph{generic computation on encrypted data}.
\autoref{fig:overview} shows how FHE enables secure offloading of computation.
The client wants to compute an expensive function $f$
(e.g., a deep learning inference) on some private data $x$.
To do this, the client encrypts $x$ and sends it to an untrusted server,
which computes $f$ on this encrypted data \emph{directly} using FHE,
and returns the encrypted result to the client.
FHE provides ideal security properties: even if the server is compromised,
attackers cannot learn anything about the data,
as it remains encrypted throughout.

\figOverview

FHE is a young but quickly developing technology.
First realized in 2009~\cite{gentry09}, early FHE schemes were
about 10$^9$ times slower than performing computations on unencrypted data.
Since then, improved FHE schemes have greatly reduced these overheads
and broadened its applicability~\cite{albrecht:hesg18:standard,peikert2016decade}.
FHE has inherent limitations---for example, data-dependent branching is impossible,
since data is encrypted---so it won't subsume all computations.
Nonetheless, important classes of computations, like deep learning inference~\cite{cheon:ictaci17:homomorphic,dathathri:pldi19:chet,dathathri:pldi20:eva},
linear algebra, and other inference and learning tasks~\cite{han:aaai19:logistic} are a good fit for FHE.
This has sparked significant industry and government investments~\cite{ibm,intel,dprive}
to widely deploy FHE.

Unfortunately, FHE still carries substantial performance overheads:
despite recent advances~\cite{dathathri:pldi19:chet, dathathri:pldi20:eva, roy:hpca19:fpga-he, brutzkus:icml19:low, polyakov:17:palisade},
FHE is still 10,000$\times$ to 100,000$\times$ slower than unencrypted computation when executed in carefully optimized software.
Though this slowdown is large, it can be addressed with hardware acceleration:
\emph{if a special- ized FHE accelerator provides large speedups over software execution,
it can bridge most of this performance gap and enable new use cases.}

For an FHE accelerator to be broadly useful, it should be programmable, i.e., capable of executing arbitrary FHE computations.
While prior work has proposed several FHE accelerators, they do not meet this goal.
Prior FHE accelerators~\cite{cousins:hpec14:fpga-he,cousins:tetc17:fpga-he,doroz:tc15:accelerating-fhe,roy:hpca19:fpga-he,riazi:asplos20:heax,turan:tc20:heaws} target individual FHE operations,
and miss important ones that they leave to software.
These designs are FPGA-based, so they are small and 
miss the data movement issues facing an FHE ASIC accelerator.
These designs also overspecialize their functional units to specific parameters,
and cannot efficiently handle the range of parameters needed within a program or across programs.

In this paper we present \name, the first programmable FHE accelerator.
\name builds on an in-depth architectural analysis of the \mbox{characteristics} of FHE computations, which exposes the main challenges and reveals the design principles a programmable FHE architecture should exploit.

\paragraph{Harnessing opportunities and challenges in FHE:}
\name is tailored to the three defining characteristics of FHE:

\noindent \textbf{\emph{(1) Complex operations on long vectors:}}
FHE encodes information using very large vectors, several thousand elements long,
and processes them using modular arithmetic.
\name employs \emph{vector processing} with \emph{wide functional units} tailored to FHE operations
to achieve large speedups.
The challenge is that two key operations on these vectors, the Number-Theoretic Transform (NTT) and automorphisms, are not element-wise and require complex dataflows that are hard to implement as vector operations. 
To tackle these challenges, \name features specialized NTT units and the first vector implementation of an automorphism functional unit.

\noindent \textbf{\emph{(2) Regular computation:}}
FHE programs are dataflow graphs of arithmetic operations on vectors.
All operations and their dependences are known ahead of time (since data is encrypted, branches
or dependences determined by runtime values are impossible).
\name exploits this by adopting \emph{static scheduling}:
in the style of Very Long Instruction Word (VLIW) processors,
all components have fixed latencies and the compiler is in charge of scheduling
operations and data movement across components, with no hardware mechanisms to handle hazards (i.e., no stall logic).
Thanks to this design, \name can issue many operations per cycle with minimal control overheads;
combined with vector processing, \name can issue tens of thousands of scalar operations per cycle. %

\noindent \textbf{\emph{(3) Challenging data movement:}}
In FHE, encrypting data increases its size (typically by at least 50$\times$);
data is grouped in long vectors; and some operations require large amounts (tens of MBs) of auxiliary data.
Thus, we find that data movement is \emph{the key challenge} for FHE acceleration:
despite requiring complex functional units, in current technology, limited on-chip storage and memory bandwidth are the bottleneck for most FHE programs.
Therefore, \name is primarily designed to minimize data movement.
First, \name features an explicitly managed on-chip memory hierarchy,
with a heavily banked scratchpad and distributed register files.
Second, \name uses mechanisms to decouple data movement and hide access latencies by loading data far ahead of its use.
Third, \name uses new, FHE-tailored scheduling algorithms that maximize reuse and make the best out of limited memory bandwidth.
Fourth, \name uses relatively \emph{few functional units with extremely high throughput}, rather than lower-throughput functional units as in prior work.
This \emph{reduces the amount of data that must reside on-chip simultaneously}, allowing higher reuse.

In summary, \name brings decades of research in architecture to bear, including vector processing and static scheduling, and combines them with new specialized functional units (\autoref{sec:FUs}) and scheduling algorithms (\autoref{sec:scheduler}) to design a programmable FHE accelerator.
We implement the main components of \name in RTL and synthesize them in a commercial 14nm/12nm process.
With a modest area budget of 151\,mm$^2$, our \name implementation 
provides 36 tera-ops/second of 32-bit modular arithmetic, 64\,MB of on-chip storage, and a 1\,TB/s high-bandwidth memory.
We evaluate \name using cycle-accurate simulation running complete FHE applications,
and demonstrate speedups of 1,200$\times$--17,000$\times$ over state-of-the-art software implementations.
These dramatic speedups counter most of FHE's overheads and enable new applications.
For example, \name executes a deep learning inference that used to take 20 minutes in 240 milliseconds,
enabling secure real-time deep learning in the cloud.

\section{Background}\label{sec:background}

Fully Homomorphic Encryption allows performing arbitrary
arithmetic on encrypted plaintext values, via appropriate operations
on their ciphertexts. Decrypting the resulting ciphertext yields the
same result as if the operations were performed on the plaintext
values ``in the clear.''

Over the last decade, prior work has proposed multiple \emph{FHE schemes},
each with somewhat different capabilities
and performance tradeoffs.
BGV~\cite{brakerski:toct14:leveled},
B/FV~\cite{brakerski:crypto12:fully,fan:iacr12:somewhat},
GSW~\cite{gentry:crypto13:homomorphic}, and CKKS~\cite{cheon:ictaci17:homomorphic} are popular FHE schemes.\footnote{These scheme names are acronyms of their authors' last names. For instance, BGV is Brakerski-Gentry-Vaikuntanathan.}~Though these schemes differ in how they encrypt plaintexts, they all
use the same data type for ciphertexts:
polynomials where each coefficient is an integer modulo $Q$.
This commonality makes it possible to build a single accelerator that supports multiple FHE schemes;
\name supports BGV, GSW, and CKKS.

We describe FHE in a layered fashion:
\autoref{sec:fhe_mapping} introduces FHE's programming model and operations, i.e., FHE's \emph{interface};
\autoref{sec:fhe_operation} describes how FHE operations are \emph{implemented};
\autoref{sec:fhe_optimizations} presents implementation \emph{optimizations};
and \autoref{sec:fhe_analysis} performs an \emph{architectural analysis}
of~a~representative FHE kernel to reveal acceleration opportunities.

For concreteness, we \emph{introduce FHE using the BGV scheme}, and briefly discuss other FHE schemes in \autoref{sec:fhe_others}.

\subsection{FHE programming model and operations}
\label{sec:fhe_mapping}

FHE programs are \emph{dataflow graphs}: directed acyclic graphs where nodes are operations and edges represent data values.
Data values are inputs, outputs, or intermediate values consumed by one or more operations.
All operations and dependences are known in advance, and data-dependent branching is impossible.

In FHE, unencrypted (plaintext) data values are always \emph{vectors};
in BGV~\cite{brakerski:toct14:leveled}, each vector consists of $N$
integers modulo an integer $t$.  BGV provides three operations on
these vectors: element-wise \emph{addition} (mod $t$), element-wise
\emph{multiplication} (mod $t$), and a small set of particular
vector \emph{permutations}.

We stress that this is BGV's \emph{interface}, not its implementation:
it describes \emph{unencrypted} data, and the homomorphic operations
that BGV implements on that data in its encrypted form.  In
\autoref{sec:fhe_operation} we describe how BGV represents encrypted
data and how each operation is implemented.

At a high level, FHE provides a vector programming model with restricted operations where 
individual vector elements cannot be directly accessed.  This causes some overheads in certain
algorithms. For example, summing up the elements of a vector is non-trivial,
and requires a sequence of permutations and additions.

Despite these limitations, prior work has devised reasonably efficient implementations of key algorithms,
including linear algebra~\cite{halevi:crypto14:algorithms},
neural network inference~\cite{brutzkus:icml19:low, gilad:icml16:cryptonets}, 
logistic regression~\cite{han:iacr18:efficient}, and genome processing~\cite{blatt:nas20:secure}.
These implementations are often coded by hand, but recent work has proposed FHE compilers to automate
this translation for particular domains, like deep learning~\cite{dathathri:pldi19:chet,dathathri:pldi20:eva}.

Finally, note that not all data must be encrypted:
BGV provides versions of addition and multiplication where one of the operands is unencrypted.
Multiplying by unencrypted data is cheaper, so algorithms can trade privacy for performance.
For example, a deep learning inference can use encrypted weights and inputs to keep the model private,
or use unencrypted weights, which does not protect the model but keeps inputs and inferences private~\cite{brutzkus:icml19:low}.

\subsection{BGV implementation overview}
\label{sec:fhe_operation}

We now describe how BGV represents and processes encrypted data (ciphertexts).
The implementation of each computation on ciphertext data is called a \emph{homomorphic operation}.
For example, the \emph{homomorphic multiplication} of two ciphertexts yields another ciphertext that,
when decrypted, is the element-wise multiplication of the encrypted plaintexts.

\paragraph{Data types:}
BGV encodes each plaintext vector as a polynomial with~$N$ coefficients
mod~$t$.
We denote the plaintext space as~$R_t$, so
\[\mathfrak{a} = a_0 + a_1x + ... + a_{N-1}x^{N-1} \in R_t\]
is a plaintext. Each plaintext is encrypted into a ciphertext consisting of two 
polynomials of~$N$ integer coefficients modulo some $Q \gg t$.
Each ciphertext polynomial is a member of~$R_Q$.

\paragraph{Encryption and decryption:}
Though encryption and decryption are performed by the client (so \name need not accelerate~them),
they are useful to understand.
In BGV, the \textit{secret key} is a polynomial $\mathfrak{s} \in R_Q$.
To encrypt a plaintext $\mathfrak{m} \in R_t$, one samples a uniformly
random $\mathfrak{a} \in R_Q$, an \emph{error} (or \emph{noise}) $\mathfrak{e} \in R_Q$ with small entries,
and computes the ciphertext $ct$ as
\begin{equation*}
  ct = (\mathfrak{a}, \mathfrak{b} = \mathfrak{a}\mathfrak{s} + t \mathfrak{e} + \mathfrak{m}).
\end{equation*}

Ciphertext $ct = (\mathfrak{a}, \mathfrak{b})$ is decrypted by
recovering
$\mathfrak{e}' = t\mathfrak{e} + \mathfrak{m} = \mathfrak{b} -
\mathfrak{a} \mathfrak{s} \bmod{Q}$, and then recovering
$\mathfrak{m} = \mathfrak{e}' \bmod t$.  Decryption is correct as long
as~$\mathfrak{e}'$ does not ``wrap around'' modulo~$Q$, i.e., its
coefficients have magnitude less than~$Q/2$.

The security of any encryption scheme relies on the ciphertexts not
revealing anything about the value of the plaintext (or the secret
key). Without adding the noise term $\mathfrak{e}$, the original message $\mathfrak{m}$ would be recoverable from $ct$ via simple Gaussian elimination.
Including the noise term entirely hides the plaintext (under cryptographic assumptions)~\cite{lyubashevsky:tact10:ideal}.

As we will see, homomorphic operations on ciphertexts increase their noise,
so we can only perform a limited number of
operations before the resulting noise becomes too
large %
and makes decryption fail.  We later describe \emph{noise management
  strategies} (Sec. \ref{noisemgmt}) %
to keep this noise bounded and thereby allow unlimited
operations.

\subsubsection{Homomorphic operations}
\paragraph{\\Homomorphic addition} of ciphertexts
$ct_0 = (\mathfrak{a}_{0}, \mathfrak{b}_{0})$ and
$ct_1 = (\mathfrak{a}_{1}, \mathfrak{b}_{1})$ is done simply by adding
their corresponding polynomials:
$ct_{\text{add}} = ct_0 + ct_1 = (\mathfrak{a}_0 + \mathfrak{a}_1,
\mathfrak{b}_0 + \mathfrak{b}_1)$.

\paragraph{Homomorphic multiplication} requires two steps.
First, the four input polynomials are multiplied and assembled:
\begin{equation*}
  ct_{\times} = (\mathfrak{l}_2, \mathfrak{l}_1, \mathfrak{l}_0) = (\mathfrak{a}_0\mathfrak{a}_1,
  \mathfrak{a}_0\mathfrak{b}_1 + \mathfrak{a}_1 \mathfrak{b}_0,
  \mathfrak{b}_0\mathfrak{b}_1) .
\end{equation*}
This $ct_{\times}$ can be seen as a special intermediate ciphertext
encrypted under a different secret key. The second step performs a
\emph{key-switch\-ing op\-era\-tion} to produce a ciphertext encrypted under
the original secret key~$\mathfrak{s}$. More specifically,
$\mathfrak{l}_2$ undergoes this key-switching process
to produce two polynomials
$(\mathfrak{u}_1, \mathfrak{u}_0) =
\textrm{KeySwitch}(\mathfrak{l}_2)$.  The final output ciphertext is
$ct_{\text{mul}} = (\mathfrak{l}_1 + \mathfrak{u}_1, \mathfrak{l}_0 +
\mathfrak{u}_0)$.

As we will see later (\autoref{sec:fhe_analysis}), key-switching is an
expensive operation that dominates the cost of a multiplication.

\paragraph{Homomorphic permutations} permute the~$N$ plaintext values
(coefficients) that are encrypted in a ciphertext.
Homomorphic permutations are implemented using \emph{automorphisms},
which are special permutations of the coefficients of the ciphertext
polynomials.  There are~$N$ automorphisms, denoted
$\sigma_k(\mathfrak{a})$ and $\sigma_{-k}(\mathfrak{a})$ for all
positive odd $k<N$. Specifically, %
\begin{equation*}
  \sigma_k(\mathfrak{a}): a_i \rightarrow (-1)^{s} a_{ik \textrm{ mod } N} \text{ for } i=0,...,N-1,
\end{equation*}
where $s=0$ if $ik \textrm{ mod } 2N < N$, and $s=1$ otherwise.
For example, $\sigma_{5}(\mathfrak{a})$ permutes $\mathfrak{a}$'s coefficients so that 
$a_0$ stays at position 0, $a_1$ goes from position 1 to position 5, and so on (these wrap around, e.g., with $N=1024$,
$a_{205}$ goes to position~1, since $205\cdot5 \textrm{ mod } 1024 = 1$).

To perform a homomorphic permutation, we first compute an automorphism on the ciphertext polynomials:
$ct_{\sigma} = (\sigma_k(\mathfrak{a}), \sigma_k(\mathfrak{b}))$.
Just as in homomorphic multiplication, $ct_{\sigma}$ is encrypted
under a different secret key, requiring an expensive key-switch to
produce the final output
$ct_{\text{perm}} = (\mathfrak{u}_1, \sigma_{k}(\mathfrak{b}) +
\mathfrak{u}_0)$, where
$(\mathfrak{u}_1, \mathfrak{u}_0) = \text{KeySwitch}(\sigma_k
(\mathfrak{a}))$.

We stress that the permutation applied to the ciphertext \emph{does
  not} induce the same permutation on the underlying plaintext
vector. For example, using a single automorphism and careful indexing,
it is possible to homomorphically \emph{rotate} the vector of the $N$
encrypted plaintext values.

\subsubsection{Noise growth and management}\label{noisemgmt}

\textrm{\\Recall} that ciphertexts have noise, which limits the number of
operations that they can undergo before decryption gives an incorrect
result.  Different operations induce different noise growth: addition
and permutations cause little growth, but multiplication incurs much
more significant growth.  So, to a first order, the amount of noise is
determined by \emph{multiplicative depth},
i.e., the longest chain of homomorphic multiplications
in the computation.

Noise forces the use of a large ciphertext modulus $Q$.
For example, an FHE program with multiplicative depth of 16
needs $Q$ to be about 512 bits.  The noise budget, and
thus the tolerable multiplicative depth, grow linearly with~$\log Q$.

FHE uses two noise management techniques in tandem:
\emph{bootstrapping} and \emph{modulus switching}.

\paragraph {Bootstrapping}~\cite{gentry09} enables FHE computations of
\emph{unbounded} depth.  Essentially, it removes noise from a
ciphertext without access to the secret key.
This is accomplished by evaluating the decryption function homomorphically.
Bootstrapping is an expensive procedure that consists of many 
(typically tens to hundreds) ho\-mo\-mor\-phic op\-era\-tions.
FHE programs with a large multiplicative depth can be divided into regions of limited depth,
separated by bootstrapping operations.

Even with bootstrapping, FHE schemes need a large noise budget
(i.e., a large~$Q$) because \emph{(1)}~bootstrapping is computationally expensive,
and a higher noise budget enables less-frequent bootstrapping, and
\emph{(2)}~bootstrapping itself consumes a certain noise budget
(this is similar to why pipelining circuits hits a performance ceiling: registers themselves add area and latency).

\paragraph{Modulus switching} rescales ciphertexts from modulus~$Q$ to a
modulus~$Q'$, which reduces the noise proportionately.
Modulus switching is usually applied before each homomorphic
multiplication, to reduce its noise blowup.

For example, to execute an FHE program of multiplicative depth 16, we
would start with a 512-bit modulus~$Q$.  Right before each multiplication, we
would switch to a modulus that is 
32 bits shorter.
So, for example, operations at depth 8 use a 256-bit
modulus.  Thus, beyond reducing noise, modulus switching reduces
ciphertext sizes, and thus computation cost.

\subsubsection{Security and parameters}
\textrm{\\The} dimension~$N$ and modulus~$Q$ cannot be chosen independently;
$N/\log Q$ must be above a certain level for sufficient security.  In
practice, this means that using a wide modulus to support deep
programs also requires a large $N$.  For example, with 512-bit $Q$,
$N=16K$ is required to provide an acceptable level of security,
resulting in very large ciphertexts.

\subsection{Algorithmic insights and optimizations}\label{sec:algoInsights}
\label{sec:fhe_optimizations}

\name leverages two optimizations developed in prior work:

\paragraph{Fast polynomial multiplication via NTTs:}
Multiplying two polynomials requires convolving their coefficients, an
expensive (naively $O(N^2)$) operation.
Just like convolutions can be made faster with the Fast Fourier Transform,
polynomial multiplication can be made faster with the Number-Theoretic Transform (NTT)~\cite{moenck1976practical},  %
a variant of the discrete Fourier transform for modular arithmetic.
The NTT takes an $N$\hyp{}coefficient polynomial as input and returns an $N$\hyp{}element vector representing the input in the
\textit{NTT domain}. Polynomial multiplication can be performed as element-wise multiplication in the NTT domain. Specifically,
\begin{equation*}
    NTT(\mathfrak{a}\mathfrak{b}) = NTT(\mathfrak{a}) \odot NTT(\mathfrak{b}),
\end{equation*}
where $\odot$ denotes component-wise multiplication. 
(For this relation to hold with $N$\hyp{}point NTTs, a \emph{negacyclic} NTT~\cite{lyubashevsky:tact10:ideal} must be used (\autoref{sec:fourStepNTT}).)

Because an NTT requires only $O(N \log N)$ modular operations, 
multiplication can be performed in $O(N \log N)$ operations by using two forward NTTs,
element-wise multiplication, and an inverse NTT.
And in fact, optimized FHE implementations often store polynomials in the NTT domain
rather than in their coefficient form \emph{across operations}, further reducing the number of NTTs.
This is possible because the NTT is a linear transformation, so additions and automorphisms can also be performed in the NTT domain:
\vspace{-0.05in} %
\begin{align*}
    NTT(\sigma_k(\mathfrak{a})) &= \sigma_k(NTT(\mathfrak{a})) \\
    NTT(\mathfrak{a} + \mathfrak{b}) &= NTT(\mathfrak{a}) + NTT(\mathfrak{b})
\end{align*}
\vspace{-0.2in}

\paragraph{Avoiding wide arithmetic via Residue Number System (RNS) representation:}
FHE requires wide ciphertext coefficients (e.g., 512 bits), but wide arithmetic is expensive:
the cost of a modular multiplier (which takes most of the compute)
grows quadratically with bit width in our range of interest.
Moreover, \mbox{we need to efficiently} %
support a broad range of widths (e.g., 64 to 512 bits in 32-bit increments),
both because programs need different widths, and because modulus switching progressively reduces coefficient widths.

RNS representation \cite{garner1959residue}  %
enables representing a single polynomial with wide coefficients as multiple polynomials with narrower coefficients,
called \emph{residue polynomials}.
To achieve this, the modulus~$Q$  is chosen to be the product of $L$
smaller distinct primes, $Q = q_1q_2\cdots\ q_L$.
Then, a polynomial in $R_Q$ can be represented as $L$ polynomials in
$R_{q_1}, \ldots, R_{q_L}$,
where the coefficients in the $i$-th polynomial are simply the wide coefficients modulo $q_i$.
For example, with $W = 32$-bit words, a ciphertext polynomial with $512$-bit modulus~$Q$ is represented as
$L = \log Q/W = 16$ polynomials with $32$-bit coefficients.

All FHE operations can be carried out under RNS representation, and have either better or equivalent bit-complexity than
  operating on one wide-coefficient polynomial.

\subsection{Architectural analysis of FHE}
\label{sec:fhe_analysis}

We now analyze a key FHE kernel in depth to understand how we can (and cannot) accelerate it.
Specifically, we consider the key-switching operation,
which is expensive and takes the majority of work in all of our benchmarks.

\autoref{listing:keyswitch} shows an implementation of key-switching.
Key\hyp{}switching takes three inputs: a polynomial \texttt{x}, and two \emph{key-switch hint matrices} \texttt{ksh0} and \texttt{ksh1}. 
\texttt{x} is stored in RNS form as $L$ residue polynomials (\texttt{RVec}). Each residue polynomial \texttt{x[i]}
is a vector of $N$ 32-bit integers modulo $q_i$.
Inputs and outputs are in the NTT domain; only the \texttt{y[i]} polynomials (line 3) are in coefficient form.

\begin{figure}
\begin{center}
  \begin{lstlisting}[caption={Key-switch implementation. \texttt{RVec} is an $N$-element vector of 32-bit values, storing a single RNS polynomial in either the coefficient or the NTT domain. %
    %
    }, mathescape=true, style=custompython, label=listing:keyswitch]
  def keySwitch(x: RVec[L], 
        ksh0: RVec[L][L], ksh1: RVec[L][L]):
    y = [INTT(x[i],$q_i$) for i in range(L)]
    u0: RVec[L] = [0, ...]
    u1: RVec[L] = [0, ...]
    for i in range(L):
      for j in range(L):
        xqj = (i == j) ? x[i] : NTT(y[i], $q_j$)
        u0[j] += xqj * ksh0[i,j] mod $q_j$
        u1[j] += xqj * ksh1[i,j] mod $q_j$
    return (u0, u1)
  \end{lstlisting}
\end{center}
\vspace{0.25cm}
\end{figure}

\paragraph{Computation vs.\ data movement:}
A single key-switch requires $L^2$ NTTs,
$2L^2$ multiplications, and $2L^2$ additions of $N$-element \mbox{vectors}.
In RNS form, the rest of a homomorphic multiplication (excluding key-switching)
is $4L$ multiplications and $3L$ additions (\autoref{sec:fhe_operation}), so key-switching is dominant.

However, the main cost at high values of $L$ and $N$ is data movement. 
For example, at $L = 16$, $N = 16K$, each RNS polynomial (\texttt{RVec}) is 64\,KB; 
each ciphertext polynomial is 1\,MB; each ciphertext is 2\,MB; and
the key-switch hints dominate, taking up 32\,MB.
With \name's compute throughput,
fetching the inputs of each key-switching from off-chip memory would demand
about 10\,TB/s of memory bandwidth.
Thus, it is crucial to reuse these values as much as possible.

Fortunately, key-switch hints can be reused:
all homomorphic multiplications use the same key-switch hint matrices,
and each automorphism has its own pair of matrices.
But values are so large that few of them fit on-chip.

Finally, note that there is no effective way to decompose or tile this operation
to reduce storage needs while achieving good reuse:
tiling the key-switch hint matrices on either dimension produces many long-lived intermediate values;
and tiling across \texttt{RVec} elements is even worse because in NTTs every input element affects every output element.

\paragraph{Performance requirements:}
We conclude that, to accommodate these large operands, an FHE accelerator requires a memory system that
\emph{(1)} decouples data movement from computation, as demand misses
during frequent key-switches would tank performance; and
\emph{(2)} implements a large amount of on-chip storage (over 32\,MB in our example)
to allow reuse across entire homomorphic operations
(e.g., reusing the same key-switch hints across many homomorphic multiplications).

Moreover, the FHE accelerator must be designed to use the memory system well.
First, scheduling data movement and computation is crucial: data must be fetched far ahead of its use to provide decoupling,
and operations must be ordered carefully to maximize reuse.
Second, since values are large, excessive parallelism can increase footprint and hinder reuse.
Thus, the system should use relatively few high-throughput functional units rather than many low-throughput ones.

\paragraph{Functionality requirements:}
Programmable FHE accelerators must support a wide range of parameters, both $N$ (polynomial/vector sizes)
and $L$ (number of RNS polynomials, i.e., number of 32-bit prime factors of $Q$). While $N$ is generally fixed for a single program,
$L$ changes as modulus switching sheds off polynomials.

Moreover, FHE accelerators must avoid overspecializing in order to support algorithmic diversity.
For instance, we have described \emph{an} implementation of key-switching, but there are others~\cite{kim:jmir18:helr,gentry:crypto2012:homomorphic}
with different tradeoffs. %
For example, an alternative implementation requires much more compute
but has key-switch hints that grow with $L$ instead of $L^2$,
so it becomes attractive for very large $L$ ($\sim$20).

\name accelerates \emph{primitive operations on large vectors}:
modular arithmetic, NTTs, and automorphisms.
It exploits wide vector processing to achieve very high throughput, even though this makes NTTs and automorphisms costlier.
\name avoids building functional units for coarser primitives, like key-switching, which would hinder algorithmic diversity.

\paragraph{Limitations of prior accelerators:}
Prior work has proposed several FHE accelerators for FPGAs~\cite{cousins:hpec14:fpga-he,cousins:tetc17:fpga-he,doroz:tc15:accelerating-fhe,roy:hpca19:fpga-he,migliore:tecs17:he-karatsuba,riazi:asplos20:heax,turan:tc20:heaws,mert:tvlsi20:bfv-accel}.
These systems have three important limitations.
First, they work by accelerating some primitives but defer others to a general-purpose host processor,
and rely on the host processor to sequence operations.
This causes excessive data movement that limits speedups.
Second, these accelerators build functional units for \emph{fixed parameters} $N$ and $L$ (or $\log Q$ for those not using RNS).
Third, many of these systems build overspecialized primitives that limit algorithmic diversity.

Most of these systems achieve limited speedups, about 10$\times$ over software baselines.
HEAX~\cite{riazi:asplos20:heax} achieves larger speedups (200$\times$ vs.\ a single core).
But it does so by overspecializing: it uses relatively low-throughput functional units for primitive operations, 
so to achieve high performance, it builds a fixed-function pipeline for key-switching.

\subsection{FHE schemes other than BGV}
\label{sec:fhe_others}

We have so far focused on BGV, but other FHE schemes provide different
tradeoffs.  For instance, whereas BGV requires integer
plaintexts, CKKS~\cite{cheon:ictaci17:homomorphic} supports ``approximate'' computation on \mbox{fixed-point} values.
B/FV~\cite{brakerski:crypto12:fully,fan:iacr12:somewhat} encodes
plaintexts in a way that makes modulus switching before 
homomorphic multiplication unnecessary, thus easing programming (but forgoing the
efficiency gains of modulo switching). And
GSW~\cite{gentry:crypto13:homomorphic} features reduced, asymmetric
noise growth under homomorphic multiplication, but encrypts a
small amount of information per ciphertext (not a full
$N/2$-element vector).

Because \name accelerates primitive operations rather than full homomorphic
operations, it supports BGV, CKKS, and GSW with the same hardware,
since they use the same primitives.  Accelerating B/FV would require
some other primitives, so, though adding support for them would not be
too difficult, our current implementation does not target it.

\section{F1 Architecture}\label{sec:arch}

\autoref{fig:arch} shows an overview of \name, which we derive from the insights in \autoref{sec:fhe_analysis}.

\paragraph{Vector processing with specialized functional units:}
\name features wide-vector execution with functional units (FUs) tailored to 
primitive FHE operations.
Specifically, \name implements vector FUs for modular addition, modular multiplication, NTTs (forward and inverse in the same unit),
and automorphisms.
Because we leverage RNS representation, these FUs use a fixed, small arithmetic word size (32 bits in our implementation),
avoiding wide arithmetic.

\figArch

FUs process vectors of configurable \emph{length} $N$ using a fixed number of \emph{vector lanes} $E$.
Our implementation uses $E=$128 lanes and supports power-of-two lengths $N$ from 1,024 to 16,384.
This covers the common range of FHE polynomial sizes, so an RNS polynomial maps to a single vector.
Larger polynomials (e.g., of 32K elements) can use multiple vectors.

All FUs are \emph{fully pipelined}, so they achieve the same throughput of $E=$128 elements/cycle.
FUs consume their inputs in contiguous chunks of $E$ elements in consecutive cycles.
This is easy for element-wise operations, but hard for NTTs and automorphisms.
\autoref{sec:FUs} details our novel FU implementations, including the first vector implementation of automorphisms.
Our evaluation shows that these FUs achieve much higher performance than those of prior work.
This is important because, as we saw in \autoref{sec:fhe_analysis},
\emph{having fewer high-throughput FUs reduces parallelism and thus memory footprint}.

\paragraph{Compute clusters:}
Functional units are grouped in \emph{compute clusters}, as \autoref{fig:arch} shows.
Each cluster features several FUs (1 NTT, 1 automorphism, 2 multipliers, and 2 adders in our implementation)
and a banked register file that can (cheaply) supply enough operands each cycle to keep all FUs busy.
The chip has multiple clusters (16 in our implementation).

\paragraph{Memory system:}
\name features an explicitly managed memory hierarchy. As \autoref{fig:arch} shows,
\name features a large, heavily banked scratchpad (64\,MB across {16} banks in our implementation).
The scratchpad interfaces with both high-bandwidth off-chip memory (HBM2 in our implementation)
and with compute clusters through an on-chip network.

\name uses decoupled data orchestration~\cite{pellauer:asplos19:buffets} to hide main memory latency.
Scratchpad banks work autonomously, fetching data from main memory far ahead of its use.
Since memory has relatively low bandwidth, off-chip data is always staged in scratchpads,
and compute clusters do not access main memory directly.

The on-chip network connecting scratchpad banks and compute clusters provides very high bandwidth,
which is necessary because register files are small and achieve limited reuse.
We implement a single-stage bit-sliced crossbar network~\cite{passas:tocaid12:crossbar} that provides full bisection bandwidth.
Banks and the network have wide ports (512 bytes), so that a single scratchpad bank can send a vector to a compute unit
at the rate it is consumed (and receive it at the rate it is produced).
This avoids long staging of vectors at the register files.

\paragraph{Static scheduling:}
Because FHE programs are completely regular, \name adopts a \emph{static, exposed microarchitecture}:
all components have fixed latencies, which are exposed to the compiler.
The compiler is responsible for scheduling operations and data transfers in the appropriate cycles to prevent
structural or data hazards.
This is in the style of VLIW processors~\cite{fisher:isca83:very}.

Static scheduling simplifies logic throughout the chip. For example, FUs need no stalling logic;
register files and scratchpad banks need no dynamic arbitration to handle conflicts;
and the on-chip network uses simple switches that change their configuration independently over time,
without the buffers and arbiters of packet-switched networks.

Because memory accesses do have a variable latency, we assume the worst-case latency,
and buffer data that arrives earlier %
(note that, because we access large chunks of data,
e.g., 64\,KB, this worst-case latency is not far from the average).

\paragraph{Distributed control:}
Though static scheduling is the hallmark of VLIW, \name's implementation is quite different:
rather than having a single stream of instructions with many operations each, in
\name each component has an \emph{independent instruction stream}.  %
This is possible because \name does not have any control flow: though FHE programs may have loops,
we unroll them to avoid all branches, and compile programs into linear sequences of instructions.

This approach may appear costly. But vectors are very long, so each instruction encodes a lot of work and this overhead is minimal. Moreover, this enables a compact instruction format, which encodes a single operation followed by the number of cycles
to wait until running the next instruction. 
This encoding avoids the low utilization of VLIW instructions, which leave many operation slots empty.
Each FU, register file, network switch, scratchpad bank, and memory controller has its own instruction stream,
which a control~unit~fetches in small blocks and distributes to components.
Overall, instruction fetches consume less than 0.1\% of memory traffic.

\paragraph{Register file (RF) design:} Each cluster in \name requires 10 read ports and 6 write ports to keep all FUs busy.
To enable this cheaply, we use an 8-banked \emph{element-partitioned} register file design~\cite{asanovic:ucb98:vector}
that leverages long vectors:
each vector is striped across banks, and each FU cycles through all banks over time, using a single bank each cycle.
By staggering the start of each vector operation, FUs access different banks each cycle.
This avoids multiporting, requires a simple RF-FU interconnect, and performs within 5\%
of an ideal infinite-ported RF.

\section{Scheduling Data and Computation}\label{sec:scheduler}

We now describe \name's software stack,
focusing on the new static scheduling algorithms
needed to use hardware well.

\figCompilerOverview

\autoref{fig:compilerOverview} shows an overview of the \name compiler.
The compiler takes as input an FHE program written in a high-level domain specific language (\autoref{sec:programming}).
The compiler is structured in three stages.
First, the \emph{homomorphic operation compiler} 
orders high-level operations to maximize reuse and
translates the program into a \emph{computation dataflow graph},
where operations are computation instructions but there are no loads or stores.
Second, the \emph{off-chip data movement scheduler} %
schedules transfers between main memory and the scratchpad to achieve decoupling and maximize reuse.
This phase uses a simplified view of hardware, considering it as
a scratchpad directly attached to functional units. %
The result is a dataflow graph that includes loads and stores from off-chip memory.
Third, the \emph{cycle-level scheduler} refines this dataflow graph.
It uses a cycle-accurate hardware model to divide instructions across compute clusters
and schedule on-chip data transfers.
This phase determine the exact cycles of all operations, and produces the instruction streams for all components.

This multi-pass scheduling primarily minimizes off-chip data movement, the critical bottleneck.
Only in the last phase do we consider on-chip placement and data movement.

\paragraph{Comparison with prior work:}
We initially tried static sched\-uling algorithms from prior work~\cite{blelloch:acm1999:provably,marchal:jpdc2019:limiting,goodman:ics1988:code,ozer:micro1998:unified,barany:odes2011:register},
which primarily target VLIW architectures.
However, we found these approaches ill-suited to \name for multiple reasons.
First, VLIW designs have less-flexible decoupling mechanisms
and minimizing data movement is secondary to maximizing compute operations per cycle.
Second, prior algorithms often focus on loops,
where the key concern is to find a compact repeating schedule,
e.g., through software pipelining~\cite{lam1989software}.
By contrast, \name has no flow control and we can 
schedule each operation independently.
Third, though prior work has proposed register-pressure-aware
instruction scheduling algorithms,
they targeted small register files and basic blocks,
whereas we must manage a large scratchpad over a much longer horizon.
Thus, the algorithms we tried either worked poorly~\cite{ozer:micro1998:unified, goodman:ics1988:code, marchal:jpdc2019:limiting} or could not scale to the sizes required~\cite{barany:odes2011:register, xu:sigplan2007:tetris, touati:ijpp2005:register, berson:pact1993:ursa}.

For example, when considering an algorithm such as Code Scheduling to Minimize Register Usage (CSR)~\cite{goodman:ics1988:code}, we find that the schedules it produces suffer from a large blowup of live intermediate values. This large footprint causes scratchpad thrashing and results in poor performance. Furthermore, CSR is also quite computationally expensive, requiring long scheduling times for our larger benchmarks. We evaluate our approach against CSR in \autoref{sec:sensitivity}.

We also attempted to frame scheduling as a register allocation problem. Effectively, the key challenge in all of our schedules is \emph{data movement}, not computation. Finding a register allocation which minimizes spilling could provide a good basis for an effective schedule. However, our scratchpad stores at least 1024 residue vectors (1024 at maximum $N = 16K$, more for smaller values of $N$), and many of our benchmarks involve hundreds of thousands of instructions, meaning that register allocation algorithms simply could not scale to our required sizes~\cite{barany:odes2011:register, xu:sigplan2007:tetris, touati:ijpp2005:register, berson:pact1993:ursa}. 

\subsection{Translating the program to a dataflow graph}
\label{sec:programming}

We implement a high-level domain-specific language (DSL) for writing \name programs.
To illustrate this DSL and provide a running example,
\autoref{listing:mv} shows the code for matrix-vector multiplication.
This follows HELib's algorithm~\cite{halevi:crypto14:algorithms}
, which \autoref{fig:MultDataflow} shows.
This toy $4 \times 16K$ matrix-vector multiply uses input ciphertexts with $N=16K$.
Because accessing individual vector elements is not possible, the code uses homomorphic rotations %
to produce each output element.

\begin{figure}
\begin{center}
  \begin{lstlisting}[caption={$(4 \times 16K)$ matrix-vector multiply in \name's DSL.}, mathescape=true, style=custompython, label=listing:mv]
p = Program(N = 16384)
M_rows = [ p.Input(L = 16) for i in range(4) ]
output = [ None for i in range(4) ]
V = p.Input(L = 16)

def innerSum(X):
  for i in range(log2(p.N)):
    X = Add(X, Rotate(X, 1 << i))
  return X

for i in range(4):
  prod = Mul(M_rows[i], V)
  output[i] = innerSum(prod)
  \end{lstlisting}
\end{center}
\vspace{0.15cm}
\end{figure}

As \autoref{listing:mv} shows, programs in this DSL are at the level
of the simple FHE interface presented in \autoref{sec:fhe_mapping}.
There is only one aspect of the FHE implementation in the DSL:
programs encode the desired noise budget ($L=16$ in our example),
as the compiler does not automate noise management.

\subsection{Compiling homomorphic operations}

The first compiler phase works at the level of the homomorphic operations
provided by the DSL. It clusters operations to improve reuse, and translates
them down to instructions.

\paragraph{Ordering} homomorphic operations seeks to maximize
the reuse of key-switch hints, which is crucial to reduce data movement (\autoref{sec:fhe_analysis}).
For instance, the program in  \autoref{listing:mv}
uses 15 different sets of key-switch hint matrices: one for the multiplies (line 12), 
and a different one for \emph{each} of the rotations (line 8).
If this program was run sequentially as written, it would cycle through all 15 key-switching hints
(which total 480\,MB, exceeding on-chip storage) four times, achieving no reuse.
Clearly, it is better to reorder the computation to perform all four multiplies, and then all four \texttt{Rotate(X, 1)}, and so on.
This reuses each key-switch hint four times.

To achieve this, this pass first clusters \emph{independent} homomorphic operations that reuse the same hint,
then orders all clusters through simple list-scheduling.
This generates schedules with good key-switch hint reuse.
\figMultDataflow

\paragraph{Translation:} Each homomorphic operation is then compiled into instructions,
using the implementation of each operation in the target FHE scheme (BGV, CKKS, or GSW).
Each homomorphic operation may translate to thousands of instructions.
These instructions are also ordered to minimize the amount of intermediates.
The end result is an instruction-level dataflow graph where every instruction
is tagged with a priority that reflects its global order.

The compiler exploits algorithmic choice.
Specifically, there are multiple implementations of key-switching (\autoref{sec:fhe_analysis}),
and the right choice depends on $L$, the amount of key-switch reuse,
and load on FUs.
The compiler leverages knowledge of operation order to estimate these and choose the right variant.

\subsection{Scheduling data transfers}
\label{sec:datatransfers}

The second compiler phase consumes an instruction-level dataflow graph and produces 
an approximate schedule that includes data transfers decoupled from computation,
minimizes off-chip data transfers, and achieves good parallelism.
This requires solving an interdependent problem: when to bring a value into the scratchpad and which one to replace
depends on the computation schedule; and to prevent stalls, the computation schedule depends on which values are in the scratchpad.
To solve this problem, this scheduler uses a simplified model of the machine:
it does not consider on-chip data movement, and simply treats all functional units 
as being directly connected to the scratchpad. %

The scheduler is greedy, scheduling one instruction at a time.
It considers instructions ready if their inputs are available in the scratchpad,
and follows instruction priority among ready ones.
To schedule loads, we assign each load a priority
\begin{equation*}
p(\text{load}) = \max \{ p(u) | u \in users(\text{load})\},
\end{equation*}
then greedily issue loads as bandwidth becomes available.
When issuing an instruction, we must ensure that there is space to store its result.
We can often replace a dead value. %
When no such value exists, we evict the value with the furthest expected time to reuse.
We estimate time to reuse as the maximum priority among unissued users of the value. 
This approximates Belady's optimal replacement policy~\cite{belady1966study}. 
Evictions of dirty data add stores to the dataflow graph.
When evicting a value, we add spill (either dirty or clean) and fill instructions to our dataflow graph. 

\subsection{Cycle-level scheduling}

Finally, the cycle-level scheduler takes in the data movement schedule produced by the previous phase,
and schedules all operations for all components considering all resource constraints and data dependences.
This phase distributes computation across clusters and manages their register files and all on-chip transfers.
Importantly, this scheduler is fully constrained by its input schedule's off-chip data movement. 
It does not add loads or stores in this stage, but it does move loads to
their earliest possible issue cycle to avoid stalls on missing operands.
All resource hazards are resolved by stalling.
In practice, we find that this separation of scheduling into data movement and instruction scheduling produces good schedules in reasonable compilation times.

This stage works by iterating through all instructions in the order produced by the previous compiler phase (\autoref{sec:datatransfers}) and determining the minimum cycle at which all required on-chip resources are available. We consider the availability of off-chip bandwidth, scratchpad space, register file space, functional units, and ports.

During this final compiler pass, we finally account for store bandwidth, scheduling stores (which result from spills) as needed. In practice, we find that this does not hurt our performance much, as stores are infrequent across most of our benchmarks due to our global schedule and replacement policy design. After the final schedule is generated, we validate it by simulating it forward to ensure that no clobbers or resource usage violations occur.

It is important to note that because our schedules are fully static, our scheduler also doubles as a performance measurement tool. As illustrated in \autoref{fig:compilerOverview}, the compiler takes in an architecture description file detailing a particular configuration of \name. This flexibility allows us to conduct design space explorations very quickly (\autoref{sec:scalability}).

\section{Functional Units}
\label{sec:FUs}

In this section, we describe \name's novel functional units.
These include the first vectorized automorphism unit (\autoref{sec:automorphism}),
the first fully-pipelined flexible NTT unit (\autoref{sec:fourStepNTT}),
and a new simplified modular multiplier adapted to FHE (\autoref{sec:modMult}).

\subsection{Automorphism unit}\label{sec:automorphism}

Because \name uses $E$ vector lanes, each residue polynomial
is stored and processed as $G$ groups, or \emph{chunks}, of $E$ elements each ($N=G\cdot E$).
An automorphism $\sigma_k$ maps the element at index $i$ to index $ki \textrm{ mod } N$;
there are $N$ automorphisms total, two for each odd $k < N$ (\autoref{sec:fhe_operation}).
The key challenge in designing an automorphism unit is that these permutations are hard to vectorize:
we would like this unit to consume and produce $E=$128 elements/cycle, but the vectors
are much longer, with $N$ up to 16\,K, and elements are permuted across different chunks.
Moreover, we must support variable $N$ \emph{and} all automorphisms.

Standard solutions fail: a 16\,K$\times$16\,K crossbar is much too large;
a scalar approach, like reading elements in sequence from an SRAM, is too slow (taking $N$ cycles);
and using banks of SRAM to increase throughput runs into frequent bank conflicts:
each automorphism ``spreads''~elements with a different stride, so regardless of the \mbox{banking} scheme,
some automorphisms will map many consecutive elements to the~same~bank.

\figAutomorphism

We contribute a new insight that makes vectorizing automorphisms simple:
if we interpret a residue polynomial as a $G \times E$ matrix,
an automorphism can always be decomposed into two independent \emph{column} and \emph{row permutations}.
If we transpose this matrix, both column and row permutations can 
be applied \emph{in chunks of $E$ elements}. \autoref{fig:automorphism} shows an example 
of how automorphism $\sigma_3$ is applied to a residue polynomial
with $N=16$ and $E=4$ elements/cycle.
Note how the permute column and row operations are local to each $4$-element chunk.
Other $\sigma_k$ induce different permutations, but with the same row/column structure.

\figautfu
Our automorphism unit, shown in \autoref{fig:aut_fu},
uses this insight to be both vectorized (consuming $E=128$ elements/cycle) and fully pipelined.
Given a residue polynomial of $N=G\cdot E$ elements, the automorphism unit first applies the column permutation
to each $E$-element input.
Then, it feeds this to a \emph{transpose unit} that reads in the whole residue polynomial interpreting it as a $G\times E$ matrix,
and produces its transpose $E\times G$.
The transpose unit outputs $E$ elements per cycle (outputting multiple rows per cycle when $G < E$).
Row permutations are applied to each $E$-element chunk, and the reverse transpose is applied.

Further, we decompose both the row and column permutations into a pipeline of sub-permutations that are \textit{fixed in hardware},
with each sub-permutation either applied or bypassed based on simple control logic; this avoids using crossbars for the $E$-element permute row and column operations.

\paragraph{Transpose unit:}
Our \textit{quadrant-swap transpose} unit transposes an $E \times E$ (e.g., $128\times 128$) matrix by recursively decomposing it into quadrants and exploiting
the identity
\begin{equation*}
  \left[ \begin{array}{c|c}
      \texttt{A} & \texttt{B}\\
      \hline
      \texttt{C} & \texttt{D}
  \end{array}\right]^{\textrm{T}} =   \left[ \begin{array}{c|c}
      \texttt{A}^{\textrm{T}} & \texttt{C}^{\textrm{T}} \\
      \hline
      \texttt{B}^{\textrm{T}} & \texttt{D}^{\textrm{T}}
  \end{array}\right].
\end{equation*}

The basic building block is a $K \times K$ \textit{quadrant-swap} unit, which swaps quadrants \texttt{B} and \texttt{C}, as shown in \autoref{fig:quadrantSwap}(left). Operationally, the quadrant swap procedure consists of three steps, each taking $K/2$ cycles:
\begin{enumerate}
  \item Cycle \texttt{i} in the first step reads \texttt{A[i]} and \texttt{C[i]} and stores them in \texttt{top[i]} and \texttt{bottom[i]}, respectively.
\item Cycle \texttt{i} in the second step reads \texttt{B[i]} and \texttt{D[i]}. The unit activates the first swap MUX and the bypass line, thus storing \texttt{D[i]} in \texttt{top[i]} and outputing \texttt{A[i]} (by reading from \texttt{top[i]}) and \texttt{B[i]} via the bypass line.
\item Cycle \texttt{i} in the third step outputs \texttt{D[i]} and \texttt{C[i]} by reading from \texttt{top[i]} and \texttt{bottom[i]}, respectively. The second swap MUX is activated so that \texttt{C[i]} is on top.
\end{enumerate}
Note that step $3$ for one input can be done in parallel with step $1$ for the next, so the unit is \emph{fully pipelined}.

\figQuadrantSwap

The transpose is implemented by a full $E \times E$ quadrant-swap followed by $\log_2E$ layers of smaller transpose units
to recursively transpose \texttt{A}, \texttt{B}, \texttt{C}, and \texttt{D}. \autoref{fig:quadrantSwap} (right) shows an implementation for $E=8$. Finally, by selectively bypassing some of the initial quadrant swaps,
this transpose unit also works for all values of $N$ ($N=G\times E$ with power-of-2 $G < E$).

Prior work has implemented transpose units for signal-processing applications,
either using registers~\cite{wang2018pipelined,zhang2020novel} or with custom SRAM designs~\cite{shang2014single}.
Our design has three advantages over prior work: it uses standard SRAM memory,
so it is dense without requiring complex custom SRAMs;
it is fully pipelined; and it works for a wide range of dimensions.

\subsection{Four-step NTT unit}\label{sec:fourStepNTT}

There are many ways to implement NTTs in hardware:
an NTT is like an FFT~\cite{cooley:moc65:algorithm}
but with a butterfly that uses modular multipliers.
We implement $N$-element NTTs (from 1K to 16K) as a composition
of smaller $E$=128-element NTTs,
since implementing a full 16K-element NTT datapath is prohibitive.
The challenge is that standard approaches result in memory access patterns
that are hard to vectorize.

\figFourStepNTT

To that end, we use the \textit{four-step variant} of the FFT algorithm~\cite{bailey:supercomputing89:FFTs},
which adds an extra multiplication to produce a vector-friendly decomposition.
\autoref{fig:fourStepNTT} illustrates 
our
four-step NTT pipeline for $E=4$;
we use the same structure with $E=128$.
The unit is fully pipelined and consumes $E$ elements per cycle.
To compute an $N=E\times E$ NTT, the unit first computes an $E$-point NTT on each $E$-element group,
multiplies each group with twiddles,
transposes the $E$ groups, and computes another $E$-element NTT on each transpose.
The same NTT unit implements the inverse NTT
by storing multiplicative factors (\textit{twiddles}) required for both forward and inverse NTTs in a small \textit{twiddle SRAM}.

Crucially, we are able to support all values of $N$ using a single four-step NTT pipeline by conditionally bypassing layers in the second NTT butterfly.
We use the same transpose unit implementation as with automorphisms.

Our four-step pipeline supports negacyclic NTTs (NCNs), which are more efficient than standard non-negacyclic NTTs (that would require padding, \autoref{sec:algoInsights}). Specifically, we extend prior work~\cite{poppelmann2015high,roy2014compact,lyubashevsky:tact10:ideal} in order to support \emph{both} forward and inverse NCNs using the same hardware as for the standard NTT. Namely, prior work shows how to either \emph{(1)} perform a forward NCN via a standard decimation-in-time (DIT) NTT pipeline, or \emph{(2)} perform an inverse NCN via a standard decimation-in-frequency (DIF) NTT pipeline. The DIF and DIT NTT variants use different hardware; therefore, this approach requires separate pipelines for forward and inverse NCNs. Prior work~\cite{lyubashevsky:tact10:ideal} has shown that separate pipelines can be avoided by adding a multiplier either before or after the NTT: doing an \emph{inverse} NCN using a \emph{DIT} NTT requires a multiplier unit \emph{after} the NTT, while doing a \emph{forward} NCN using a \emph{DIF} NTT requires a multiplier unit \emph{before} the NTT.

We now show that \emph{both} the forward and inverse NCN can be done in the same standard four-step NTT pipeline, with \emph{no additional hardware}. This is because the four-step NTT already has a multiplier and two NTTs in its pipeline. We set the first NTT to be decimation-in-time and the second to be decimation-in-frequency (\autoref{fig:fourStepNTT}). To do a forward NTT, we use the forward NCN implementation via DIT NTT for the first NTT; we modify the contents of the Twiddle SRAM so that the multiplier does the pre-multiplication necessary to implement a forward NCN in the second NTT (which is DIF and thus requires the pre-multiplication). Conversely, to do an inverse NTT, we modify the Twiddle SRAM contents to do the post\hyp{}mul\-ti\-pli\-ca\-tion necessary to implement an inverse NCN in the first NTT (which is DIT); and we use the inverse NCN imple\-men\-ta\-tion via DIF NTT for the second NTT.

The NTT unit is large: each of the 128-element NTTs requires $E(\log (E)-1)/2$=384 multipliers,
and the full unit uses 896 multipliers.
But its high throughput improves performance over many low-throughput NTTs (\autoref{sec:evaluation}). %
This is the first implementation of a fully-pipelined four-step NTT unit, 
improving NTT performance by 1,600$\times$ over the state of the art (\autoref{sec:perf}).

\subsection{Optimized modular multiplier}\label{sec:modMult}
\tblModMult

Modular multiplication computes $a\cdot b \textrm{ mod } q$.
This is the most expensive and frequent operation.
Therefore, improvements to the modular multiplier have an almost
linear impact on the computational capabilities of an FHE accelerator.

Prior work~\cite{mert:euromicro19:design}
recognized that a Montgomery multiplier~\cite{montgomery:mom85:modular} within NTTs can be improved by leveraging
the fact that the possible values of modulus $q$ are restricted by the number of elements the NTT is applied to.
We notice that if we only select moduli $q_i$, such that $q_i = -1 \textrm{ mod } 2^{16}$,
we can remove a mutliplier stage from~\cite{mert:euromicro19:design};
this reduces area by 19\% and power by 30\% (\autoref{tbl:modMult}).
The additional restriction on $q$ is acceptable because FHE
requires at most 10s of moduli~\cite{gentry:crypto2012:homomorphic},
and our approach allows for 6,186~prime~moduli.

\section{\name Implementation}
\label{sec:implementation}

We have implemented \name's components in RTL,
and synthesize them in a commercial 14/12nm process using state-of-the-art tools.
These include a commercial SRAM compiler that we use for scratchpad and register file banks.

We use a dual-frequency design: most components run at 1\,GHz,
but memories (register files and scratchpads)
run double-pumped at 2\,GHz.
Memories meet this frequency easily and this enables using single-ported SRAMs while serving up to two accesses per cycle.
By keeping most of the logic at 1\,GHz, we achieve higher energy efficiency.
We explored several non-blocking on-chip networks (Clos, Benes, and crossbars).
We use 3 16$\times$16 bit-sliced crossbars~\cite{passas:tocaid12:crossbar} (scratch\-pad$\rightarrow$cluster, cluster$\rightarrow$scratchpad, and cluster$\rightarrow$cluster). %

\autoref{tbl:GF12} shows a breakdown of area by component, as well as the area of our \name configuration,
151.4\,mm$^2$.
FUs take 42\% of the area, with 31.7\% going to memory,
6.6\% to the on-chip network, and 19.7\% to the two HBM2 PHYs.
We assume 512\,GB/s bandwidth per PHY;
this is similar to the NVIDIA A100 GPU~\cite{choquette2021nvidia}, which has 2.4\,TB/s with 6 HBM2E PHYs~\cite{nvidiadgx}.
We use prior work to estimate HBM2 PHY area~\cite{rambuswhite, dasgupta20208} and power~\cite{rambuswhite, ge2011design}.

This design is constrained by memory bandwidth: though it has 1\,TB/s of bandwidth,
the on-chip network's bandwidth is 24\,TB/s, and the aggregate bandwidth between RFs and FUs is 128\,TB/s.
This is why maximizing reuse is crucial.

\section{Experimental Methodology}

\paragraph{Modeled system:}
We evaluate our \name implementation from \autoref{sec:implementation}.
We use a cycle-accurate simulator to execute \name programs.
Because the architecture is static, this is very different from conventional simulators,
and acts more as a checker: it runs the instruction stream at each component and verifies
that latencies are as expected and there are no missed dependences or structural hazards.
We use activity-level energies from RTL synthesis to produce energy breakdowns.

\paragraph{Benchmarks:}
We use several FHE programs to evaluate \name. %
All programs come from state-of-the-art software implementations, which we port to \name:

\subparagraph{Logistic regression}
uses the HELR algorithm~\cite{han:aaai19:logistic}, which is based on CKKS.
We compute a single batch of logistic regression training with up to $256$ features, and $256$ samples per batch,
starting at computational depth $L = 16$; this is equivalent to the first batch of HELR's MNIST workload. 
This computation features %
ciphertexts with large $\log Q$ ($L = 14,15,16$), so it needs careful data orchestration to run efficiently.

\subparagraph{Neural network} benchmarks come from Low Latency CryptoNets (LoLa)~\cite{brutzkus:icml19:low}.
This work uses B/FV, an FHE scheme that \name does not support, so we use CKKS instead.
We run two neural networks:
LoLa-MNIST is a simple, LeNet-style network used on the MNIST dataset~\cite{lecunn:ieee98:gradient-document},
while LoLa-CIFAR is a much larger 6-layer network (similar in computation to MobileNet v3~\cite{howard2019searching})
used on the CIFAR-10 dataset~\cite{cifar10}.
LoLa-MNIST includes two variants with unencrypted and encrypted weights;
LoLa-CIFAR is available only with unencrypted weights.
These three benchmarks use relatively low $L$ values (their starting $L$ values are 4, 6, and 8, respectively),
so they are less memory-bound.
They also feature frequent automorphisms,
showing the need for a fast automorphism~unit.

\tblGF %

\subparagraph{DB Lookup} is adapted from HELib's \texttt{BGV\_country\_db\_lookup}~\cite{helib:db-lookup}. A BGV-encrypted query string is used to traverse an encrypted key-value store and return the corresponding value. The original implementation uses a low security level for speed of demonstration, but in our version, we implement it at $L=$17, $N=$16K for realism. We also parallelize the CPU version so it can effectively use all available cores. DB Lookup is both deep and wide, so running it on \name incurs substantial off-chip data movement.

\addtocounter{table}{1}
\tblMicrobenchmark

\subparagraph{Bootstrapping:} We evaluate bootstrapping benchmarks for BGV and CKKS.
Bootstrapping takes an $L=1$ ciphertext with an exhausted noise budget and refreshes it
by bringing it up to a chosen top value of $L=L_{max}$, then performing the bootstrapping computation
to eventually obtain a usable ciphertext at a lower depth (e.g., $L_{max} - 15$ for BGV).

For BGV, we use Sheriff and Peikert's algorithm~\cite{alperin:crypto13:practical} for non-packed BGV boot\-strap\-ping, with $L_{max} = 24$.
This is a particularly challenging benchmark because it features computations at large values of $L$.
This exercises the scheduler's 
algorithmic choice component, which selects
the right key-switch method to balance computation and data movement.

For CKKS, we use non-packed CKKS bootstrapping from HEA\-AN~\cite{cheon:eurocrypt2018:bootstrapping}, also with $L_{max} = 24$.
CKKS bootstrapping has many fewer ciphertext multiplications than BGV, greatly reducing
reuse opportunities for key-switch hints.

\paragraph{Baseline systems:}
We compare \name with a CPU system running the baseline programs (a 4-core, 8-thread, 3.5\,GHz Xeon E3-1240v5).
Since prior accelerators do not support full programs, we also include microbenchmarks of single operations
and compare against HEAX~\cite{riazi:asplos20:heax}, the fastest prior accelerator.

\section{Evaluation}\label{sec:evaluation}
\subsection{Performance}\label{sec:perf}

\addtocounter{table}{-2}
\tblBenchmark
\addtocounter{table}{1}

\paragraph{Benchmarks:}
\autoref{tbl:benchmark} compares the performance of \name and the CPU on full benchmarks.
It reports execution time in milliseconds for each program (lower is better), and \name's speedup over the CPU (higher is better).
\name achieves dramatic speedups, from 1,195$\times$ to 17,412$\times$ (5,432$\times$ gmean).
CKKS bootstrapping has the lowest speedups as it's highly memory-bound;
other speedups are within a relatively narrow band, as compute and memory traffic are more balanced.

These speedups greatly expand the applicability of FHE. Consider deep learning:
in software, even the simple LoLa-MNIST network takes seconds per inference,
and a single inference on the more realistic LoLa-CIFAR network takes \emph{20 minutes}.
\name brings this down to 241 \emph{milliseconds},
making real-time deep learning inference practical:
when offloading inferences to a server, this time is comparable
to the roundtrip latency between server and client.

\paragraph{Microbenchmarks:}
\autoref{tbl:microbenchmark} compares the performance of \name, the CPU, and HEAX$_\sigma$ on four microbenchmarks:
the basic NTT and automorphism operations on a single ciphertext,
and homomorphic multiplication and permutation (which uses automorphisms).
We report three typical sets of parameters.
We use microbenchmarks to compare against prior accelerators,
in particular HEAX.
But prior accelerators do not implement automorphisms,
so we extend each HEAX key-switching pipeline with an SRAM-based, scalar automorphism unit.
We call this extension HEAX$_\sigma$.

\autoref{tbl:microbenchmark} shows that
\name achieves large speedups over HEAX$_\sigma$,
ranging from 172$\times$ to 1,866\x.
Moreover, \name's speedups over the CPU are even larger than in full benchmarks.
This is because microbenchmarks are pure compute,
and thus miss the data movement bottlenecks of FHE programs.

\subsection{Architectural analysis}

To gain more insights into these results, we now analyze \name's data movement, power consumption, and compute.

\paragraph{Data movement:}
\autoref{fig:dataMovement} shows a breakdown of off-chip memory traffic across data types:
key-switch hints (KSH), inputs/outputs, and intermediate values.
KSH and input/output traffic is broken into compulsory and
non-compulsory (i.e., caused by limited scratchpad capacity).
Intermediates, which are always non-compulsory, are classified as loads or stores.

\autoref{fig:dataMovement} shows that key-switch hints dominate in high-depth workloads
(LogReg, DB Lookup, and bootstrapping), taking up to 94\% of traffic.
Key-switch hints are also significant in the LoLa-MNIST variants.
This shows why scheduling should prioritize them.
Second, due our scheduler design, \name approaches compulsory
traffic for most benchmarks, with non\hyp{}compulsory accesses
adding only 5-18\% of traffic.
The exception is LoLa-CIFAR, where intermediates consume 75\% of traffic.
LoLa-CIFAR has very high reuse of key-switch hints,
and exploiting it requires spilling intermediate ciphertexts.

\figDataMovement
\figOpBreakdown

\paragraph{Power consumption:}
\autoref{fig:power} reports average power for each benchmark, broken down by component.
This breakdown also includes off-chip memory power (\autoref{tbl:GF12} only included the on-chip component).
Results show reasonable power consumption for an accelerator card.
Overall, computation consumes 20-30\% of power, and data movement dominates.

\paragraph{Utilization over time:}
\name's average FU utilization is about 30\%.
However, this doesn't mean that fewer FUs could achieve the same performance:
benchmarks have memory\hyp{}bound phases 
that weigh down average FU utilization.
To see this, \autoref{fig:opBreakdown} shows a breakdown of FU utilization over 
time for LoLa-MNIST Plaintext Weights.
\autoref{fig:opBreakdown} also shows off-chip bandwidth utilization over time (black line).
The program is initially memory-bound, and few FUs are active.
As the memory-bound phase ends, compute intensity grows, 
utilizing a balanced mix of the available FUs.
Finally, due to decoupled execution,
when memory bandwidth utilization peaks again,
\name can maintain high compute intensity.
The highest FU utilization happens at the end of the benchmark and is caused by processing
the final (fully connected) layer, which is highly parallel and already has all inputs available on-chip.

\subsection{Sensitivity studies}
\label{sec:sensitivity}

\tblSensitivity

To understand the impact of our FUs and scheduling algorithms, we evaluate \name variants without them.
\autoref{tbl:sensitivity} reports the \emph{slowdown (higher is worse)} of \name with:
\emph{(1)} low\hyp{}throughput NTT FUs that follow the same design as HEAX
(processing one stage of NTT butterflies per cycle); %
\emph{(2)} low\hyp{}throughput automorphism FUs using a serial SRAM memory,
and \emph{(3)} Goodman's register-pressure-aware scheduler~\cite{goodman:ics1988:code}.

For the FU experiments, our goal is to show the importance of having high-throughput units.
Therefore, the low-throughput variants use many more (NTT or automorphism) FUs,
so that aggregate throughput across all FUs in the system is the same.
Also, the scheduler accounts for the characteristics of these FUs.
In both cases, performance drops substantially, by gmean 2.6$\times$ and 3.3$\times$.
This is because achieving high throughput requires excessive parallelism,
which hinders data movement, forcing the scheduler to balance both.

Finally, the scheduler experiment uses register-pressure-aware scheduling~\cite{goodman:ics1988:code}
as the off-chip data movement scheduler instead, operating on the full dataflow graph.
This algorithm was proposed for VLIW processors and register files; we apply it to the larger scratchpad.
The large slowdowns show that prior capacity-aware schedulers are ineffective on \name.

\figConfigs

\subsection{Scalability}
\label{sec:scalability}

Finally, we study how \name's performance changes with its area budget: 
we sweep the number of compute clusters, scratchpad banks, HBM controllers,
and network topology to find the most efficient design at each area.
\autoref{fig:pareto} shows this
Pareto frontier, with
area in the $x$-axis and performance in the $y$-axis.
This curve shows that, as \name scales, it uses resources efficiently:
performance grows about linearly through a large range of areas.

\subsection{Functional Simulation}\label{sec:functional_simulation}

Here we describe our software simulation efforts 
for F1. Currently, we have a functional simulator
written in C++ on top of Shoup's Number Theory
Library.\footnote{\url{https://libntl.org/}}
This simulator measures \emph{input-output correctness}
and \emph{calls to functional units} throughout a computation.
The underlying algorithms are not the same
as F1's functional units, but they match
common methods used in software (i.e., HElib's algorithms). This allows one
to verify correctness of FHE algorithms and to create a
dataflow graph.
The simulator has all our functional
units implemented in software: modular additions, modular
multiplications, automorphisms, and NTTs.
We then build ciphertext-level operations by
calls to these algorithms: ciphertext addition,
ciphertext multiplication, rotations, modulus-switching,
and a simplified bootstrapping procedure, for non-packed
ciphertexts.
Our functional simulator works for the parameter
ranges discussed throughout the paper: 
polynomial/ring dimension $N$ as an arbitrary 
power of 2 (usually 1024-16384 for security)
and RNS moduli where each is an NTT-friendly 
prime, $q_i \equiv 1 \bmod 2N$, roughly
24 bits long. Further, each moduli is sampled
randomly, similarly to other FHE RNS implementations.

\section{Related Work}
\label{sec:related}

We now discuss related work not covered so far.

\paragraph{FHE accelerators:}
Prior work has proposed accelerators for individual FHE operations, but not full FHE computations~\cite{cousins:hpec12:sipher-fpga,cousins:hpec14:fpga-he,cousins:tetc17:fpga-he,doroz:tc15:accelerating-fhe,roy:hpca19:fpga-he,mert:tvlsi20:bfv-accel,migliore:tecs17:he-karatsuba,riazi:asplos20:heax,turan:tc20:heaws}.
These designs target FPGAs and rely on a host processor;
\autoref{sec:fhe_analysis} discussed their limitations.
Early designs accelerated small primitives like NTTs, and were dominated by host-FPGA communication.
State-of-the-art accelerators execute a full homomorphic multiplication independently:
Roy et al.~\cite{roy:hpca19:fpga-he} accelerate B/FV multiplication by 13$\times$ over a CPU;
HEAWS~\cite{turan:tc20:heaws} accelerates B/FV multiplication, and uses it to speed a simple 
benchmark by 5$\times$;
and HEAX~\cite{riazi:asplos20:heax} accelerates CKKS multiplication and key-switching by up to 200$\times$.
These designs suffer high data movement (e.g., HEAX does not reuse key-switch hints)
and use fixed pipelines with relatively low-throughput FUs.

We have shown that accelerating FHE programs requires a different approach:
data movement becomes the key constraint, requiring new techniques
to extract reuse {across} homomorphic operations;
and fixed pipelines cannot support the operations of even a single benchmark.
Instead, \name achieves flexibility and high performance by exploiting
wide-vector execution with high-throughput FUs.
This lets \name execute not only full applications, but different FHE schemes.

\paragraph{Hybrid HE-MPC accelerators:}
Recent work has also proposed ASIC accelerators for some homomorphic encryption primitives
in the context of oblivious neural networks~\cite{juvekar2018gazelle,reagen:hpca21:cheetah}.
These approaches are very different from FHE:
they combine homomorphic encryption with multi-party computation (MPC),
executing a single layer of the network at a time and sending intermediates
to the client, which computes the final activations.
Gazelle~\cite{juvekar2018gazelle} is a low-power ASIC for homomorphic evaluations,
and Cheetah~\cite{reagen:hpca21:cheetah} introduces algorithmic optimizations
and a large ASIC design that achieves very large speedups over Gazelle.

These schemes avoid high-depth FHE programs, so server\hyp{}side homomorphic operations are cheaper.
But they are limited by client-side computation and client-server communication:
Cheetah and Gazelle
use cipher\-texts that are up to $\sim40\times$ small\-er than those used by \name; 
however, they re\-quire the client to re\--en\-crypt
ciphertexts \textit{every} time they are mul\-ti\-pli\-ed on the server to prevent noise blowup.
CHOCO~\cite{vanderhagen:arxiv21:choco} shows that client-side computation costs for HE-MPC are substantial,
and when they are accelerated, network latency and throughput overheads dominate
(several seconds per DNN inference).
By contrast, \name enables offloading the full inference using FHE,
avoiding frequent communication. As a result, a direct comparison between these accelerators and F1 is not possible.

F1's hardware also differs substantially from Cheetah and Gazelle.
First, Cheetah and Gazelle implement fixed-function pipelines (e.g., for output-stationary DNN inference in Cheetah),
whereas F1 is programmable.
Second, Cheetah, like HEAX, uses many FUs with relatively low throughput,
whereas F1 uses few high-throughput units (e.g., 40$\times$ faster NTTs).
Cheetah's approach makes sense for their small ciphertexts,
but as we have seen (\autoref{sec:sensitivity}), it is impractical for FHE.

\paragraph{GPU acceleration:}
Finally, prior work has also used GPUs to accelerate different FHE schemes,
including GH~\cite{wang:hpec12:fhe-gpu,wang:tc13:fhe-gpu}, BGV~\cite{wang:iscas14:leveled-gpu},
and B/FV~\cite{al:emerging19:implementation}.
Though GPUs have plentiful compute and bandwidth,
they lack modular arithmetic, their pure data-parallel approach
makes non-element-wise operations like NTTs expensive,
and their small on-chip storage adds data movement.
As a result, GPUs achieve only modest performance gains.
For instance, Badawi et al.~\cite{al:emerging19:implementation}
accelerate B/FV multiplication using GPUs, and achieve speedups of around 10$\times$ to 100$\times$
over single-thread CPU execution
(and thus commensurately lower speedups over multicore CPUs, as FHE operations parallelize well).

\section{Conclusion}

FHE has the potential to enable computation offloading with guaranteed security.
But FHE's high computation overheads currently limit its applicability to narrow
cases (simple computations where privacy is paramount).
\name tackles this challenge, accelerating full FHE computations by over 3-4 orders of magnitude.
This enables new use cases for FHE, like secure real-time deep learning inference.

\name is the first FHE accelerator that is programmable,
i.e., capable of executing full FHE programs.
In contrast to prior accelerators, which build fixed pipelines tailored to specific FHE schemes and parameters,
\name introduces a more effective design approach:
it accelerates the \emph{primitive} computations shared by higher-level operations
using novel high\hyp{}throughput functional units,
and hardware and compiler are co-designed to minimize data movement,
the key bottleneck. %
This flexibility makes \name broadly useful:
the same hardware can accelerate all operations within a program,
arbitrary FHE programs, and even multiple FHE schemes.
In short, our key contribution is to show that, for FHE,
we can achieve ASIC-level performance without sacrificing programmability.

\section*{Acknowledgments}

We thank the anonymous reviewers, Maleen Abeydeera, Hyun Ryong Lee, Quan Nguyen, Yifan Yang, Victor Ying, Guowei Zhang, and Joel Emer for feedback on the paper; Tutu Ajayi, Austin Rovinski, and Peter Li for help with the HDL toolchain setup; Shai Halevi, Wei Dai, Olli Saarikivi, and Madan Musuvathi for email correspondence.
This research was developed with funding from the Defense Advanced Research Projects Agency (DARPA) under contract number Contract No. HR0011-21-C-0035. The views, opinions and/or findings expressed are those of the author and should not be interpreted as representing the official views or policies of the Department of Defense or the U.S. Government.
Nikola Samardzic was supported by the Jae S. and Kyuho Lim Graduate Fellowship at MIT.

\bibliographystyle{IEEEtranS}

\end{document}